\documentclass[
reprint,
prb,
superscriptaddress,
nofootinbib,
nobibnotes,
amsmath,amssymb
]{revtex4-2}

\usepackage{graphicx}
\usepackage{dcolumn}
\usepackage{bm}
\usepackage[colorlinks,allcolors=blue]{hyperref}

\newcommand{\op}[1]{\hat{\mathcal{#1}}}

\begin{document}

\preprint{arXiv:2503.05655}

\title{Memory preservation and cooperative shielding in complex quantum networks}

\author{Simone Ausilio}
\affiliation{Dipartimento di Fisica “Enrico Fermi”, Universit\`a di Pisa, Pisa, Italy}
\affiliation{University of Birmingham, Edgbaston, Birmingham, UK}

\author{Fausto Borgonovi}
\affiliation{Dipartimento di Matematica e Fisica and I-LAMP, Universit\`a Cattolica del Sacro Cuore, Brescia, Italy}
\affiliation{INFN, sezione di Milano, Italy}

\author{Giuseppe Luca Celardo}
\affiliation{Dipartimento di Fisica e Astronomia, Universit\`a di Firenze}
\affiliation{INFN, sezione di Firenze, Italy}

\author{Jorge Yago Malo}
\email[Corresponding author: ]{jyagom@gmail.com}
\author{Maria Luisa Chiofalo}
\affiliation{Dipartimento di Fisica “Enrico Fermi”, Universit\`a di Pisa, Pisa, Italy}
\affiliation{INFN, sezione di Pisa, Italy}

\date{\today}

\begin{abstract}
Complex quantum networks are powerful tools in the modeling of transport phenomena, particularly for biological systems, and enable the study of emergent phenomena in many-body quantum systems. High connectivity and long-range interactions induce strong constraints on the system dynamics. Here, we study the transport properties of a quantum network described by the paradigmatic XXZ Hamiltonian, with non-trivial graph connectivity and topology, and long-range interactions. We show how long-range interactions induce memory preserving effects and strongly affect the spreading of the excitations due to cooperative shielding.
We describe the  memory-preserving effect in all-to-all connected regular networks with distance-independent couplings. Indeed, the memory of the number of initially injected excitations is preserved over long times, encoded in the number of frequencies present in the dynamics. Interestingly, we find that memory-preserving effects occur also in less regular graphs, such as quantum networks with either power-law node connectivity or complex, small-world type, architectures. 
We discuss the implications of these properties in biology-related problems, such as an application to Weber's law in neuroscience, and their implementation in specific quantum technologies via biomimicry. We also show how the presence of long-range interaction strongly affects the dynamics of the excitations in small-world networks and power law all-to-all coupled networks. 
Indeed, because of cooperative shielding blue, as the connectivity or the range of interaction increases, the initial excitation spreads more slowly among the network and becomes strongly dependent on the initial conditions.
\end{abstract}

\maketitle


\section{Introduction}
\label{sec:intro}

Dynamical processes in quantum networks are relevant for the engineering of quantum simulators and quantum computing platforms~\cite{Daley2024}, for the investigation of biological media under driven-dissipative conditions~\cite{plenio2008dephasing,Chisholm_2021,Civolani2024}, and to map complex, non-linear phenomena that do not require to be quantum via the novel approach called quantum-like paradigm~\cite{YagoMalo2023,khrennikov2010paradigm,basieva2020paradigm}. Complex quantum networks are also becoming powerful tools for extracting relevant information from many-body quantum systems~\cite{LCarr}, such as emergent entanglement structures, topological instabilities, and self-similarity in quantum states, which may not be apparent by traditional methods~\cite{Cirac:2010,Nokkala:2016,VParigi}.

The importance of quantum networks, regular or complex, is enhanced by their accurate and precise engineering in current experimental setups, constituting a predominant area of development for quantum simulators~\cite{georgescu2014simulation,Altman2024,yagomalo2024review}, with widely transversal applications since the first proposals for quantum analog models more than 20 years ago~\cite{Daley2024}. Importantly, the current degree of experimental tunability and microscopic understanding of these platforms, both in their coherent and dissipative couplings \cite{daley2014qt}, has enabled the quantum simulation of models well beyond condensed matter physics~\cite{hofstetter2018condensed} or quantum computing, ranging from fundamental physics and cosmology~\cite{Bauer2023}, metrology and sensing~\cite{pezze2018metro} or -- most relevant to this work -- quantum thermalization and chaos~\cite{eisert2010entropy},\cite{Borgonovi2016} quantum chemistry~\cite{cao2019chemistry,Bauer2020} and biology~\cite{cao2020biology}.

Indeed, the characterization of quantum networks dynamics under open quantum system conditions constitutes a paradigmatic testbed for understanding information processing and transport phenomena in complex media both quantum and classical~\cite{biamonte2019complexnetworks}, addressing fundamental questions in diverse contexts. These include quantum information theory~\cite{swingle2018otoc} with transversal implications for models of quantum gravity~\cite{bhattacharyya2022quantum, bentsen2019fast}; material science, in parallel with the experimental and technological development of solid-state nanodevices and cold atom technologies~\cite{datta2005transport,Lambert2021}; biological media, where the role of coherent processes in out-of-equilibrium phenomena, though still debated~\cite{Fleming2011}, has been proven in general complex networks and biological examples \cite{albert2002complexnetworks, ball2011biology, huelga2013vibrations, plenio2008dephasing, vattay2014quantum, biamonte2019complexnetworks}. In fact, in more recent years the study of dynamics in complex networks topologies evolving under paradigmatic Hamiltonians, has raised attention to effectively investigate chemistry and biology applications, by different strategies that can be classified under two main categories. One is a multiscale approach that combines the quantum-chemistry computation of microscopic interaction potentials with the quantum simulation of energy or excitation transport in quantum networks~\cite{cao2019chemistry, cao2020biology, ball2011biology}. Another one consists in mapping the complex (classical) network into a quantum network, and exploiting the underlying superposition and entanglement at work to describe the system nonlinearities~\cite{yagomalo2024review}.

The quantum transport properties of a given system are governed by a set of general properties, all of which can be implemented in current quantum technology platforms: (i) the connectivity graph or network topology~\cite{Kulvelis2015,mulken2016complexnetworks}, where the role of quantum interference produces drastic differences between classical and quantum regimes~\cite{biamonte2019complexnetworks}; (ii) the presence of external couplings in the form of driving~\cite{Kohler2005,Alamo2024}, disorder~\cite{Moix_2013,Walschaers2016,Zhou2016} or dissipation~\cite{plenio2008dephasing,Wu_2022}, and (iii) the coherent Hamiltonian parameters governing the static properties of the model, i.e. not just the transfer rates and interaction strengths, but also the range of such couplings~\cite{Kloss2020,Dominguez2021,Defenu2023} with the latter being one of our focuses. Technological development, particularly in ion traps~\cite{Bruzewicz2019} and cold gases~\cite{Ritsch2013, Adams_2020, Chomaz_2023}, has allowed to explore conditions where individual quantum systems are accurately prepared to couple over large distances. The long-range nature of the interactions has been studied in detail and it is known to affect the static and dynamical properties of a quantum model~\cite{Defenu2023}. As to the static properties, it has been revealed to enrich the interplay between length scales in the system~\cite{Botet:1982,Liu:2019,Defenu:2018}, leading to non-additive energy contributions~\cite{Campa2009}, novel phases of matter, and distinguished changes in  spin systems~\cite{Maghrebi:2017,Giachetti_2021,Defenu2023} or cold-atom systems~\cite{Landig:2016,Habibian:2013,Sharma:2022} phase diagrams. As to the dynamical behavior~\cite{Campa2009,Dauxois2002}, long-range interactions lead to a diversity of interesting phenomena. These concern the occurrence of non-thermal topological steady states \cite{Borgonovi2004,Borgonovi2005}, excitation propagation beyond the Lieb-Robinson picture~\cite{lieb1972bound, Schachenmayer2013, Eisert2013, Hauke2013}, new behavior in the presence of disorder~\cite{Chen2023,chavez2021}, dissipation~\cite{Catalano2023}, applications to quantum reservoir computing~\cite{sannia2024_skineqnn} or, essential for the current work, trapped dynamics due to interference in what has been denoted as cooperative shielding \cite{santos2016cooperative,celardo2016_shielding}.

This phenomenon, which will be relevant in the following, appears in systems with long-range interaction between its constituents. It refers to the existence of protected subspaces within which the evolution remains unaffected for a finite time due to the appearance of approximate selection rules, with the dynamics heavily depending on the initial state. To better understand what is meant by shielding, we exemplify the concept for a simple situation. Let us consider a Hamiltonian $\op{H}=\op{H}_1+\op{H}_2$ for a many-body system, having commuting parts $[\op{H}_1,\op{H}_2]=0$. An initial state $|\psi_0\rangle$ belonging to a (degenerate or non-degenerate) eigenspace of $\op{H}_2$ and with corresponding eigenvalue $\lambda_2$, evolves in time according to the general Hamiltonian $|\psi(t)\rangle=\text{exp}(-\mathrm{i}\lambda_2t)\text{exp}(-\mathrm{i}\op{H}_1t)|\psi_0\rangle$: since $\op{H}_2$ solely induces a global phase without other effects, the dynamics is fully determined by $\op{H}_1$. So in this case we can say that the dynamics is shielded from the effects of $\op{H}_2$. On the contrary, if the initial state is disseminated across more than one eigenspace of $\op{H}_2$ its dynamics is characterized by $\op{H}$.

The concept of shielding can be extended to the case of non-commuting Hamiltonian parts in the event of $\op{H}_2$ being composed of solely long-range interactions. If the initial state is initialized in a selected subspace of $\op{H}_2$, long-range interactions do not influence its temporal evolution for a time span that scales with the system size. Instead, the evolution of an initial state with components in more than one subspace, is affected by long-range interactions: the dynamics is not bounded, with components in various subspaces exhibiting fast propagation due to the long-range coupling. Cooperative shielding can induce an effective localization of the initial excitation in long-range interacting system. This effect is counter-intuitive since one could expect that the larger the range of the interaction is, the easier the initial excitation will spread away from its initial position. Here we show that this effect has a considerable impact on the dynamics of highly connected networks. 

The cooperative shielding effect shows how  long-range interactions not always lead to a fast spreading of information, but, depending on initial conditions, the spreading of information  can remain local, thus preserving the memory of the initial state for long times and affecting the transport properties of the system. Indeed, long-range interactions have revealed to be crucial in preserving over long time both temporal and spatial memory of excitation quenches in a regular, all-to-all connected XXZ open quantum network: this model resulted from the quantum mapping of a classical neuronal complex system and demonstrated~\cite{YagoMalo2023} to effectively describe the sense of number~\cite{burr2008visual,stoianov2012vision}, the very general ability of humans and many animals of non-cognitively counting with constant precision of about 20\% up to a few hundred items in a given space or time region~\cite{burr2008visual}. Given the paradigmatic nature of the XXZ model ~\cite{franchini2017integrable}, this time-space memory preserving feature can be predicted to apply to a variety of physical systems, besides being relevant for the development of quantum technology devices.

Understanding the physical mechanism governing this
memory-preserving property is therefore of primary importance, especially when considering the whole variability of transport conditions represented by (i)-(iii) above. To this aim, analytical descriptions are particularly valuable in combination with information that can be extracted from quantum simulating the time and spectral signals of the network dynamical observables. 

In this work we focus on two related but not identical aspects: the memory preserving effect and the transport properties of different networks in presence of long-range interactions. We first provide a complete analytical understanding of the memory effect  in regular, all-to-all connected XXZ quantum networks, in which symmetries play a pivotal role. Then, we turn to the aforementioned points (i) and (iii) to parametrize the extent to which this phenomenon survives in more general cases. Specifically, we analyze  complex quantum networks described by the XXZ Hamiltonian having couplings based on a small-world architecture; moreover, we move back to completely connected quantum networks with power-law varying interactions. In both cases, cooperative shielding  plays a fundamental role in the memory preservation effect and in the transport properties of the system.
This systematic study sets the stage for a deeper exploration of information and transport processing in paradigmatic XXZ quantum (complex) networks, and the possible presence of universal behavior, revealing some degree of order in disordered structures.
The article is organized as follows. In Section~\ref{sec:symm_and_dyn}, we introduce the XXZ model, analyze its symmetry structure, and exploit the latter to identify the physical mechanism for time-space memory preservation of excitations. The analytics are performed by first studying the all-to-all connected regular XX network and then adding the $z$-axis anisotropy interaction perturbatively.
In Section~\ref{sec:trans_gen_net}, we study both transport, via the analysis of the Inverse Participation Ratio, and the memory effect in highly-connected random networks and in all-to-all connected regular networks with power-law decaying interactions.
Finally, in Section~\ref{sec:conclusion}, we discuss the relevance of our findings for the phenomenological description of complex systems, and the perspective research questions that our work opens up.

\section{Symmetry and dynamics in all-to-all connected quantum networks}
\label{sec:symm_and_dyn}

\begin{figure*}[!tb]
    \centering \includegraphics[width=0.95\linewidth]{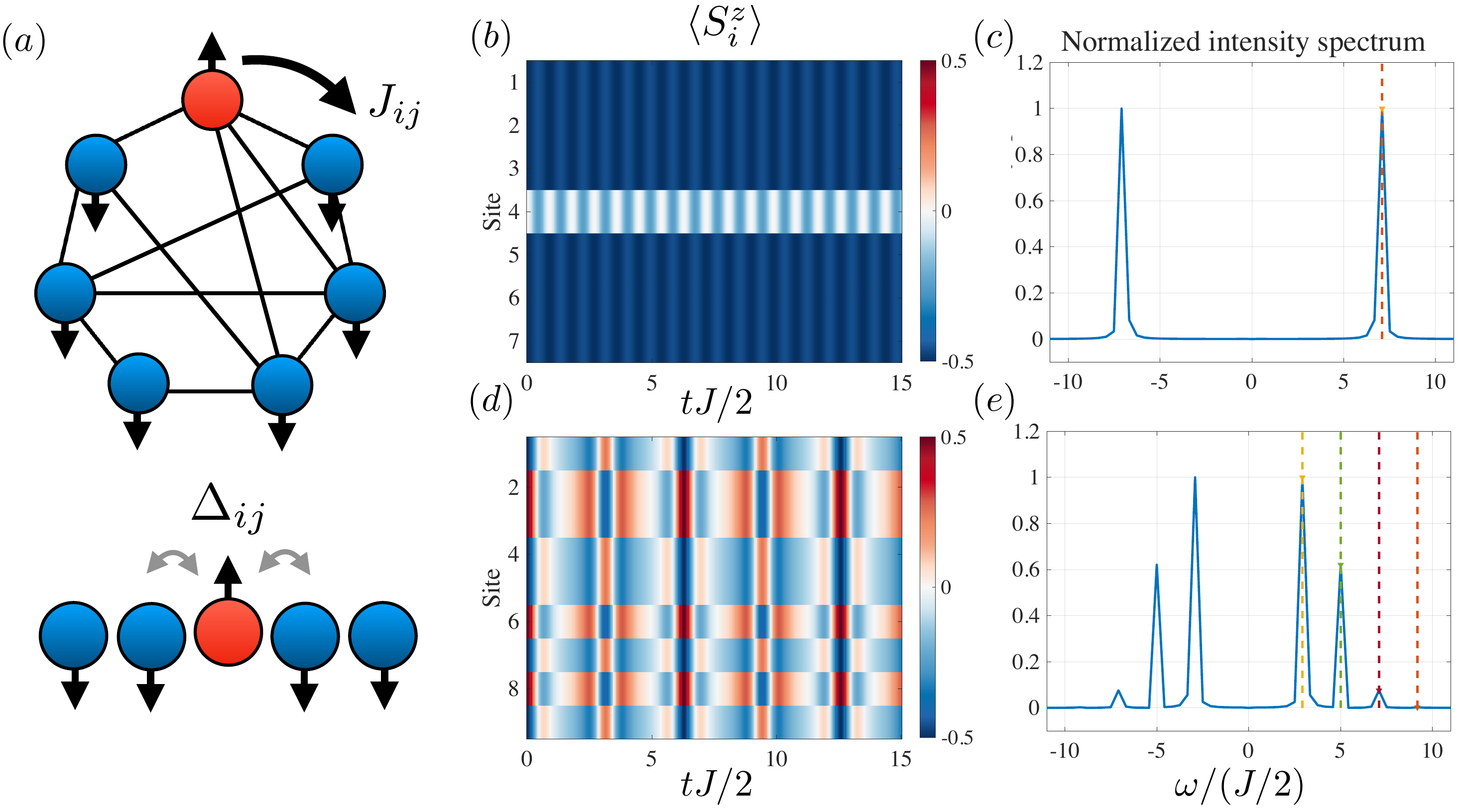}
    \caption{The role of symmetry on  the quantum network dynamics. $(a)$ Quantum spin-1/2 XXZ network. Excitations along the $z$-axis are depicted by red circles (spin up) in a background of non-excited blue circles (spin downs). The system presents two-body interactions by the spin-exchange coupling $J_{ij}$ and the anisotropy $\Delta_{ij}$. $(b)-(e)$ Temporal evolution of the local $z$-magnetization for the all-to-all connected XXZ spin network, revealing how the network preserves time and space memory of the excitations introduced over time. $(b)$ refers to a network with $L=7$ sites and one excitation quench positioned at site 4 along $x$, i.e. with initial state $|\phi(0)\rangle=(|\downarrow\downarrow\downarrow\uparrow\downarrow\downarrow\downarrow\rangle+|\downarrow\downarrow\downarrow\downarrow\downarrow\downarrow\downarrow\rangle)/\sqrt{2}$; the arrows represent the spin direction along the $z$-axis. Note that the signal at site 4 starts from a value of 0, instead of +0.5, because $\langle\phi(0)|\op{S}_4^z|\phi(0)\rangle$=0. $(d)$ refers to a network with $L=9$ sites and four excitations quenches along $z$ positioned at $i=2, 3, 6, 8$, thus the initial state is $|\downarrow\uparrow\uparrow\downarrow\downarrow\uparrow\downarrow\uparrow\downarrow\rangle$. Example parameters chosen are $\Delta_{ij}$=0 and $J_{ij}/2=1$, d$t$=0.06 (units of $J/2$) and Periodic Boundary Conditions (PBC). Plots $(c)$ and $(e)$ depict the normalized power spectra of $\langle \op{S}_i^z\rangle$ in site $i=4$ for cases $(b)$ and $(d)$ respectively, revealing how the number of peaks precisely reflects the number of excitations in the system. The vertical dashed red line corresponds to the possible frequency of the dynamics as computed in  Section~\ref{sec:xx-ham}. The data has been detrended before the Fourier transform by removing the zero-frequency component.}
    \label{fig:state-art}
\end{figure*}

We begin by introducing the paradigmatic Heisenberg XXZ Hamiltonian we are interested in, with its relevant properties and symmetries; throughout the paper we will work in natural units, i.e. $\hbar=1$. As schematized in Fig.~\ref{fig:state-art}$(a)$, the XXZ Hamiltonian is given by:
\begin{equation}
\begin{split}
    \op{H} & =\sum_{(i,j)}\Bigl[J_{ij}(\op{S}_i^x\op{S}_j^x+\op{S}_i^y\op{S}_j^y)+\Delta_{ij}\op{S}_i^z\op{S}_j^z\Bigr] \\
    & =\sum_{(i,j)}\Bigl[\frac{J_{ij}}{2}(\op{S}_i^+\op{S}_j^-+\op{S}_i^-\op{S}_j^+)+\Delta_{ij}\op{S}_i^z\op{S}_j^z\Bigr]\,,
\end{split}
\label{hamiltonian1}
\end{equation}
where $\op{S}_i^{x,y,z}$ represents the local spin-1/2 magnetization on site $i$ along the given direction and $\op{S}_i^{\pm}$ represent the corresponding ladder operators for the local spin-1/2 Hilbert space. The sum runs over all the edges $(i,j)$ of a given graph $G$, representing the interactions between sites. Therefore, for each site the physical neighbors are given by the vertex neighbors on the graph (see Fig.~\ref{fig:state-art}$(a)$). Quasi-particles are excited in the system, in the form of spin states (e.g. up, red circles) in a background of opposite spins (e.g. down, blue circles); for simplicity, throughout the paper we will call such quasi-particles \textit{excitations}. The number of excitations indicate the number of spins up (along the $z$ direction) in a background of spins down. Excitations can hop from site to site   via exchange interaction $J_{ij}$ between spins $i$ and $j$. The last term $\Delta_{ij}$ represents an additional interaction between the $z$-components of the spins. 

The XXZ model has been studied at length for regular graphs with remarkable differences in local and global observables depending on the degree of coupling~\cite{YagoMalo2023, yagomalo2024review, santos2016cooperative, franchini2017integrable}. Here, we focus on the case of highly-connected networks. We also focus on  the important example of all-to-all exchange coupling (where all the $J_{ij}$ are the same independently of the distance) for different initial conditions, see  Fig.~\ref{fig:state-art}$(b)$ Fig.~\ref{fig:state-art}$(d)$. The initial condition is taken to be an eigenstate of the total $z$-magnetization operator $\op{M}_z=\sum_i\op{S}_i^z$, which assume the general form $|\boldsymbol{\sigma}_q\rangle=|\sigma_1^{q}\dots\sigma_L^{q}\rangle$, with each $\sigma$ denoting either spin-up, $\uparrow$, or spin-down, $\downarrow$; this string will be dubbed as \textit{Pauli string}, while the basis set $\{|\boldsymbol{\sigma}_q\rangle, q=1,\dots,2^L\}$ will be called \textit{computational basis}.

In the all-to-all connected case, the interference pattern that emerges from the high connectivity leads to substantial differences in the evolution, even at long times, compared to the familiar local models. From a physical point of view, when the excitation is located at the initial site $i$, it has equal probability to tunnel to any other node if $\Delta=0$. This leads to a regular and symmetric evolution as it is seen in the power spectrum of the time signal, Fig.~\ref{fig:state-art}$(c)$ and $(e)$, generated by a standard Fast Fourier Transform (FFT) applied to  the time evolution of the expectation value of local $z$-magnetization operators $\langle \op{S}_i^z\rangle$; note that this analysis is independent of the system's dimensionality. Notably, the power spectrum  shows a  number of discrete frequencies equal to  the  number of initial excitations, see Section~\ref{sec:xx-ham}. Notice that when considering several excitations, while the time signal profile appears complex (Fig.~\ref{fig:state-art}$(d)$), the power spectrum is still well-defined (Fig.~\ref{fig:state-art}$(e)$).

The role of symmetries in highly-connected graphs, Fig.~\ref{fig:state-art}$(a)$, for integrable models has paramount importance to understand their quantum transport properties. We now analyze in detail the consequences of the Hamiltonian symmetries on the energy levels and their effects on the dynamics. To this aim, we begin with the simplest case of the XX Hamiltonian, i.e. $\Delta_{ij}=0$ in Eq.\eqref{hamiltonian1}, and then we turn it on again as a perturbation. 

\subsection{Eigenspaces of the exchange (XX) Hamiltonian}\label{sec:xx-ham}
We start by analyzing only the exchange part of the Hamiltonian of Eq.\eqref{hamiltonian1} with all-to-all interactions, and considering it as our unperturbed mean-field component:
\begin{equation*}
    \op{H}_0=J\sum_{i<j}(\op{S}_i^x\op{S}_j^x+\op{S}_i^y\op{S}_j^y)=\frac{J}{2}\sum_{i<j}(\op{S}_i^+\op{S}_j^-+\op{S}_i^-\op{S}_j^+)\,.
\end{equation*}
We show below that $\op{H}_0$ can be rewritten as function of the total $z$-magnetization $\op{M}_z=\sum_i\op{S}_i^z$ and of the (Casimir invariant) total squared magnetization $\op{M}^2=\op{M}_x^2+\op{M}_y^2+\op{M}_z^2$, see \cite{santos2016cooperative}. We then re-cast the operators $\op{S}_i^x\op{S}_j^x$ (equivalent for the $y$-component) in terms of  $\op{M}_z$ and  $\op{M}^2$ as:
\begin{eqnarray*}
\op{M}_x^2=\sum_k(\op{S}_k^x)^2+2\sum_{i<j}\op{S}_i^x\op{S}_j^x\, \\\,\sum_{i<j}\op{S}_i^x\op{S}_j^x=\frac{\op{M}_x^2-\sum_k(\op{S}_k^x)^2}{2}\,.
\end{eqnarray*}
Additionally, we express each local spin operator as $(\op{S}_k^x)^2+(\op{S}_k^y)^2=\op{S}_k^2-(\op{S}_k^z)^2=1/2(1/2+1)-(1/2)^2=1/2$, the action of these operators on the single particles being known. Thus, the Hamiltonian becomes:
\begin{equation}
    \op{H}_0=\frac{J}{4}\Bigl[2(\op{M}^2-\op{M}_z^2)-L\Bigr],
    \label{hamiltonian30}
\end{equation}
with $L$ the system size (number of spins).
The result in Eq.\eqref{hamiltonian30} implies that although the starting Hamiltonian is the sum of two-body interacting terms, the all-to-all connectivity allows for the appearance of collective behavior, where the system behaves as if it had a single, large spin \cite{omanakuttan2023scrambling}.
Labeling the quantum numbers associated to operators $\op{M}^2$ and $\op{M}_z$ with $l$ and $m$, respectively, we use the angular momenta composition rules for spin-$1/2$ to calculate the corresponding energy bands $E_{lm}$:
\begin{align}
    l & =\begin{cases}
    0,1,\dots,\frac{L}{2}\quad\text{if $L$ is even} \\
    \frac{1}{2},\frac{3}{2},\dots,\frac{L}{2}\quad\text{if $L$ is odd} \\
    \end{cases} \notag \\
    m & =-l,(-l+1),\dots,(l-1),l \notag \\
    E_{lm} & =\frac{J}{4}\Bigl[2\Bigl(l(l+1)-m^2\Bigr)-L\Bigr]\,.
    \label{energ_lev}
\end{align}
Degeneracy-wise, we can compute the number $f(l, L)$ of different ways one can get a certain value of l for a given $L$, through the following recursive formula:
\[
\begin{cases}
    f\Bigl(\frac{1}{2}, 1\Bigr)=1 \\
    f(l\text{ not allowed}, L)=0 \\
    f(l, L)=f\Bigl(l-\frac{1}{2}, L-1\Bigr)+f\Bigl(l+\frac{1}{2}, L-1\Bigr)
\end{cases}
.
\]
The Hamiltonian $\mathbb{Z}_2$ symmetry also implies that $E_{l,m}=E_{l,-m}$. Hence, the level degeneracy is:
\begin{equation*}
    \text{deg}(E_{l|m|}, L)=\begin{cases}
    2f(l, L)\quad |m|\neq 0 \\
    f(l, L)\quad |m|=0 \\
    \end{cases}\,.
\end{equation*}
We give a proof and details of the computation in Appendix~\ref{appdx:degeneracy_formula}.

\begin{figure}[!tb]
    \centering    \includegraphics[width=\linewidth]{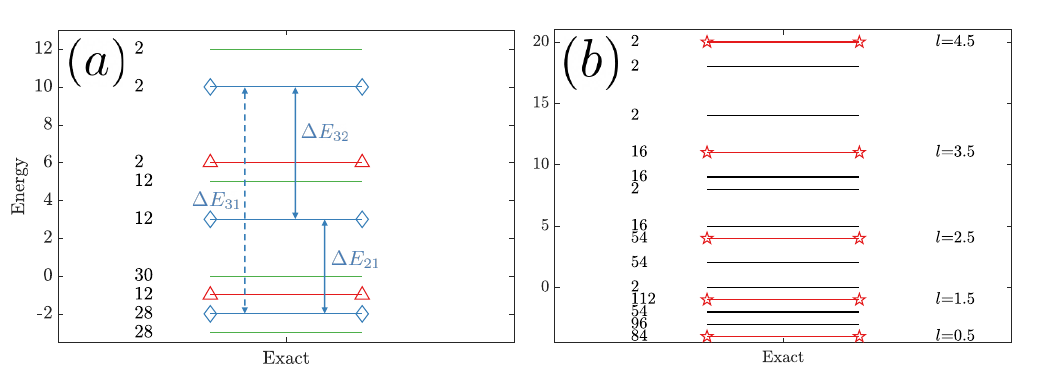}    \caption{Symmetries and structure of the energy levels for the all-to-all connected XX Hamiltonian. $(a)$ Diagram for a system with $L=7$ sites and constant exchange coupling $J/2=1$ (unperturbed study). The vertical axis represents the exact values of the eigenvalues of $\op{H}_0$, while the numbers on the left of the horizontal lines account for the degeneracy of the corresponding level; different colors and symbols represent different subspaces corresponding to different number of excitations along the $z$-axis. Red triangles and lines: levels corresponding to  1 excitation. Blue lozenges and lines: levels referring to 2 excitations. Double-headed arrows : the three possible energy gaps described by Eq.\eqref{2exc_freq}: the dotted arrow represents the frequency that is not manifested in the $z$-magnetization spectrum. Green bars: remaining energy levels. $(b)$ The same diagram for the case $L=9$, with the red bars contoured by stars representing the levels with 4 excitations  denoted by the quantum number $l$ on the right.}
    \label{energy_levels}
\end{figure}

We depict in Fig.~\ref{energy_levels} the energies and degeneracies of the all-to-all graph, highlighting the energy levels belonging to the subspaces that correspond to one (triangles) and two (diamonds) spins up. They spread across two and three energy bands, respectively, as it is dictated by the possible choices of the quantum number $l$.

We are now in a position to explain the local $z$-magnetization spectra in Fig.~\ref{fig:state-art} in terms of the energy levels. Due to the $U(1)$ symmetry, the quantum number $m$ will be fixed during the time evolution. If we consider the case of one single excitation to start with, we have $m_1=(-L/2)+1$, hence only the values $l_1=(L/2)-1$ and $l_2=L/2$ are allowed, the corresponding bands having energies $E_{l_1m_1} =-{J}/2\doteq E_1^{(1)}$ with degeneracy $L-1$ and $E_{l_2m_1}  ={J(L-1)}/2\doteq E_2^{(1)}$ with degeneracy 1. Thus, the oscillation of the local $z$-magnetization signal occurs at angular frequency equal to the energy split between the two levels:

\begin{equation}
    \Delta E=E_2^{(1)}-E_1^{(1)}=\frac{JL}{2}\,.
    \label{freq_spliten}
\end{equation}

We notice that the local $z$-magnetization spectra depend explicitly on the separation between the energy levels of the system, see vertical dashed line in Fig.~\ref{fig:state-art}$(c)$.

We can use the same line of reasoning for the second case highlighted in Fig.~\ref{energy_levels}$(a)$, with two excitations. Here, the quantum number $m$ is fixed to be $m_2=(-L/2)+2$, implying that $l$ can take values $l_1=(L/2)-2$, $l_2=(L/2)-1$ and $l_3=L/2$, reflecting in the energy levels as $E_{l_1m_2}  =-J\doteq E_1^{(2)}$ with degeneracy $L(L-3)/2$; $E_{l_2m_2}  ={J(L-4)}/2\doteq E_2^{(2)}\,\text{with degeneracy $L-1$;}$ $E_{l_3m_2}  =J(L-2)\doteq E_3^{(3)}\,\text{ with degeneracy 1.}$
The resulting energy gaps are:
\begin{align}
    \Delta E_{21}^{(2)} & =E_2^{(2)}-E_1^{(2)}=\frac{J(L-2)}{2} \notag \\
    \Delta E_{32}^{(2)} & =E_3^{(2)}-E_2^{(2)}=\frac{JL}{2} \notag \\
    \Delta E_{31}^{(2)} & =E_3^{(2)}-E_1^{(2)}=\Delta E_{32}^{(2)}+\Delta E_{21}^{(2)}=J(L-1)\,.
    \label{2exc_freq}
\end{align}

\begin{figure*}[!tb]
    \centering
    \includegraphics[width=\textwidth]{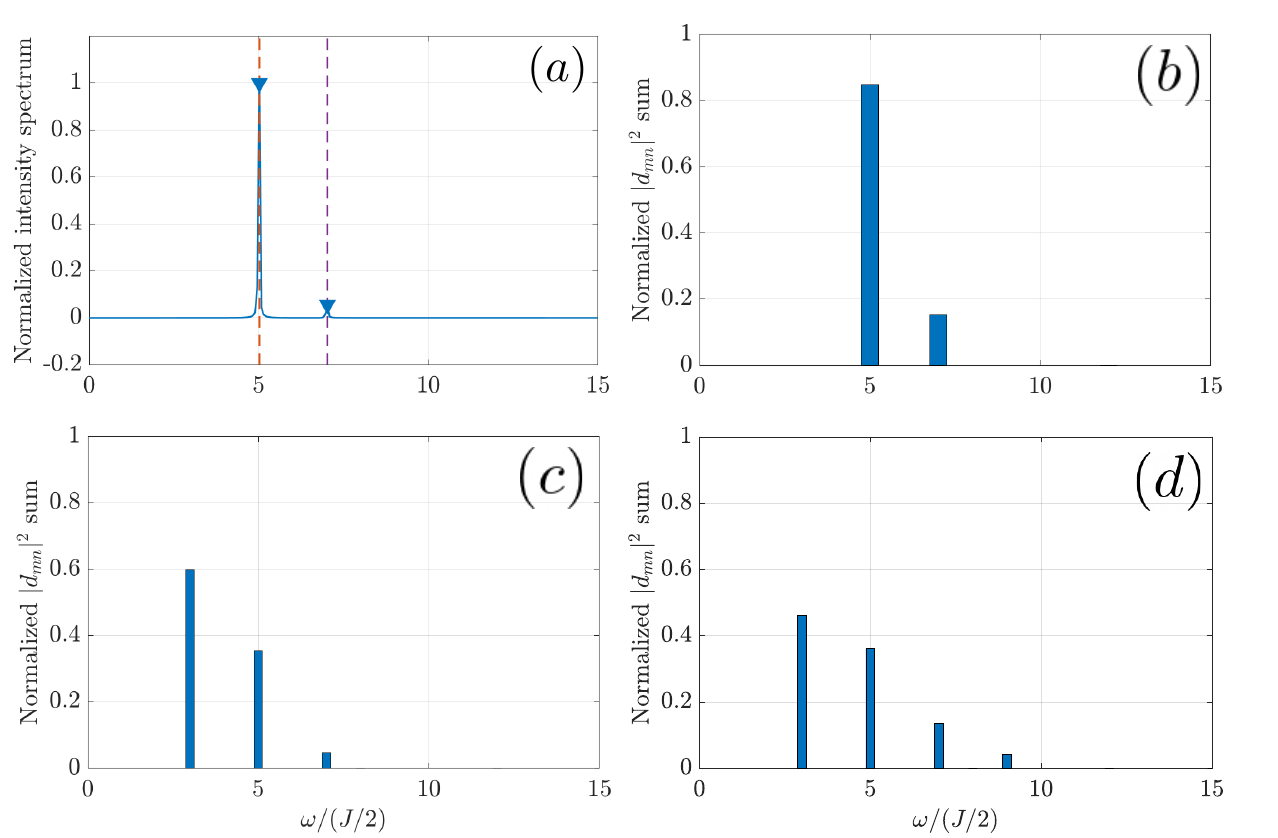}
    \caption{Multiple excitations with all-to-all connected XX network: frequency analysis. $(a)$ Spectrum's peaks for the local $z$-magnetization signal in the case of two excitations: the corresponding positions are 5 (higher prominence) and 7 (lower prominence). The system parameters are the same as in Fig.~\ref{fig:state-art}, with the signal being picked up from site 4. $(b)-(d)$ Bar plot representing the distribution of the coefficients $|d_{mn}|^2$  in Eq.\eqref{d_coeff}, showing that the number of frequencies perfectly matches with the simulated spectra in Fig.~\ref{fig:state-art}. As in the case of the spectra, the zero-frequency is not shown to highlight the oscillating components of the signals. The coefficients are summed over the levels' degeneracies and normalized to the total sum of the coefficients. $(b)$ $L=7$ and two excitations. $(c)$ $L=7$ and three excitations. $(d)$ $L=9$ and four excitations.}
    \label{twoexc_barplot}
\end{figure*}

Not all the possible energy differences appear in the power spectrum of the dynamics. To show this, we analyze in Fig.~\ref{twoexc_barplot} the power spectrum of the local magnetization $\langle\op{S}_i^z\rangle(t)$. For instance, in the case of two excitations, see Fig.~\ref{twoexc_barplot}(a), we observe the presence of two frequencies, corresponding to the gaps $\Delta E_{21}$ and $\Delta E_{32}$, while the third possible transition is absent. Even in the case depicted in Fig.~\ref{fig:state-art}$(d)-(e)$, corresponding to four initial excitations or $m=-L/2+4$, we ascertain the presence of four out of the $4\cdot(4-1)/2=6$ possible frequencies arising from the five levels having quantum number $l=L/2, L/2-1, L/2-2, L/2-3, L/2-4$; again, these four frequencies are exactly those coming from the difference between contiguous levels. The schematics for this situation is shown in Fig.~\ref{energy_levels}$(b)$.

To understand why it is so, we consider that for any expectation value of an operator as function of time and with initial state $|\psi_0\rangle$ we can write:
\begin{eqnarray}
    \langle\op{O}\rangle(t) && =\langle\psi_0|e^{\mathrm{i}\op{H}t}\op{O}e^{-\mathrm{i}\op{H}t}|\psi_0\rangle \nonumber \\
    && =\sum_{m,n}\underbrace{\langle\psi_0|\phi_m\rangle\langle\phi_n|\psi_0\rangle\langle\phi_m|\op{O}|\phi_n\rangle}_{d_{mn}} e^{-\mathrm{i}\Delta E_{mn}t},
    \label{d_coeff}
\end{eqnarray}
where we have decomposed the initial state $|\psi_0\rangle$ temporal evolution in the Hamiltonian eigenbasis $\{|\phi_n\rangle\}$, with corresponding eigenvalues $\{E_n\}$ (the pedices here represent generic quantum numbers). In our specific case, the eigenstates of the system Hamiltonian $\op{H}_0$ are those of the angular momentum $|\alpha;l,m\rangle$, with $l$ and $m$ defined in Eq.\eqref{energ_lev} and $\alpha$ another quantum number independent of angular momentum, the initial state is a superposition of Pauli strings, and the operator $\op{O}$ is a local spin operator $\op{S}^z_i\doteq\op{O}_0^{(1)}$ at site $i$: this operator has the property of being the $q=0$ component of a spherical tensor of rank $r=1$, the other components being $\op{O}_{+1}^{(1)}=-c\op{S}^+_i$ and $\op{O}_{-1}^{(1)}=c\op{S}^-_i$ with $c$ a positive constant. The coefficients $\langle\psi_0|\phi_m\rangle=\langle\psi_0|\alpha;l,m\rangle$ and $\langle\phi_n|\psi_0\rangle=\langle\beta;l',m'|\psi_0\rangle$ are the Clebsch-Gordan coefficients between the eigenbasis and the computational basis. The matrix elements of $\op{O}_0^{(1)}$ can be expressed thanks to the Wigner-Eckart theorem \cite{sakurai1994} as:

\begin{eqnarray*}
    \langle\phi_m|\op{O}|\phi_n\rangle && =\langle\alpha;l,m|\op{O}_0^{(1)}|\beta;l',m'\rangle \\
    && \propto \langle l,m|r=1,q=0;l',m'\rangle\langle\alpha;l||\op{\mathbf{O}}^{(1)}||\beta;l'\rangle,
\end{eqnarray*}
with the first term being a Clebsch-Gordan coefficient and the second the reduced matrix element for tensor $\op{\mathbf{O}}^{(1)}=\Bigl(\op{O}_{+1}^{(1)},\op{O}_{0}^{(1)},\op{O}_{-1}^{(1)}\Bigr)$. Put it another way, the theorem states that operating with a rank-$r$ spherical tensor with states possessing an angular momentum $l$ is equivalent to studying the composition of two angular momenta, one with spin $r$ and the other with spin $l$. This analogy allows to define selection rules for the Clebsch-Gordan coefficients, which for the case at study translates in:
\begin{equation}
    \begin{cases}
    |l-r|\le l'\le l+r \Rightarrow \Delta l=0, \pm 1\\
    m'=m+q \Rightarrow m'=m \\
    \end{cases}\\
    \label{selec-rules}
\end{equation}
In other words, the system symmetries (notably invariance by rotations) guarantee that the connected matrix elements are those separated by $\Delta l\leq1$ (and same value of $m$).

This property leads to the following key result: given an initial state with $b$ excitations, i.e. $b$ spin-up spins oriented along $z$ and so total magnetization $m=-L/2+b$; this initial state has support exclusively on the $b+1$ eigenspaces $l=L/2, L/2-1,\dots,L/2-b$, thus, due the selection rules in Eq.~\eqref{selec-rules}, we can observe exactly $b$ frequencies in the power spectrum of $\langle\op{S}^z_i\rangle$.

Furthermore, the selection rule $m'=m$, valid for $q=0$, allows to distinguish between the different subspaces of the total $z$-magnetization $\op{M}_z$ in case where the initial state is a superposition of states belonging to diverse magnetization sectors, like in Fig.~\ref{fig:state-art}$(b)$. Considering this example, the initial state lives among the levels $(l=L/2,m=-L/2), (l=L/2,m=-L/2+1), (l=L/2-1,m=-L/2+1)$; the described selection rule isolates the total $z$-magnetization sector between them, so that the frequencies originated from the mechanism above for $m=-L/2+1$ do not mix up with those arising from the subspace with $m=-L/2$, with this remaining true for more general situations.

As a further numerical demonstration, we dub $d_{mn}$ the coefficients depending on the eigenbasis overlaps with the initial state and the operator's matrix elements, see Eq.~\eqref{d_coeff}. The weights on the different exponentials are given by the sums of the corresponding $|d_{mn}|^2$ coefficients. We show in Fig.~\ref{twoexc_barplot}(b) the numerical evaluation of the coefficients normalized to their total sum. We notice that the  position of the two resulting finite frequencies perfectly matches that displayed in the numerically simulated  power spectrum, Fig.~\ref{twoexc_barplot}(a). In particular, we observe a zero weight for $\Delta E_{31}$, in agreement with the absence of this frequency in the simulated spectrum. While the value of the frequencies depends on the system size $L$, we show in Fig.~\ref{twoexc_barplot}(c) that their number does not. In fact, the number of frequencies only depends on the number of initial excitations, as it is visible in Fig.~\ref{twoexc_barplot}(d) for the case of 3 initial excitations, where one additional finite frequency adds on.

This analytical result embodies the physical interpretation of the simulational findings from \cite{YagoMalo2023, yagomalo2024review}: the all-to-all connected quantum network is capable of preserving long-lasting memory in space and time of the number of initially injected excitations, this memory being encoded in frequency spectrum of the collective spin excitations.

To summarize, our analytical treatment evidences how the local $z$-magnetization spectra reveal the accessible gaps between consecutive energy levels, which in turn determine the number of excitations involved in the system. The accessible gaps are traced back to the Hamiltonian symmetry and their size enhanced by the all-to-all coupling condition, thereby favoring the protection of system subspaces according to the cooperative shielding concept. In simpler words, this outcome is linked to the collective behavior of the fully-connected spin network with constant couplings, thus generating an effective, single large spin that blends together the single sites. \\
With this understanding at hand, we now proceed to analyze  the effect of the anisotropy $z$-term in Hamiltonian \eqref{hamiltonian1}.

\subsection{Role of the anisotropy (ZZ-term interaction)}
Once the energy bands of the exchange Hamiltonian $\op{H}_0$ are computed,  we can consider the effect of the anisotropy/interaction term in the Hamiltonian~\ref{hamiltonian1}:
\begin{equation}
    \op{W}=\sum_{i<j}\Delta_{ij}\op{S}_i^z\op{S}_j^z\,.
    \label{pert}
\end{equation}
To this aim, we resort to perturbation theory and calculate to first order the energy shifts due to this contribution in Eq.~\eqref{pert}. 

Since the levels are degenerate, we should in principle compute (at first order) the eigenvalues of the $\op{W}$ perturbation restricted to the considered eigenspace. However, it can be easily checked by inspection that $\op{W}$ is already expressed in a diagonal form, due to the definitions of the computational basis $\mathcal{P}$. Therefore, we only need a restricted number of matrix elements. For any form of the $\Delta_{ij}$ coefficients one can cast the action of the perturbation operator on a given state in the computational basis $|\boldsymbol{\sigma}_q\rangle$, with $\boldsymbol{\sigma}_q$ expressing the spin polarizations vector in each site, the index $q$ denoting the specific Pauli string. We thus obtain:
\begin{equation}
    \langle\boldsymbol{\sigma}_q|\op{W}|\boldsymbol{\sigma}_q\rangle=\sum_{i<j}\Delta_{ij}\frac{\text{sgn}(i,q)\text{sgn}(j,q)}{4}\,,
    \label{pert_general}
\end{equation}
where:
\begin{align*}
    \text{sgn}(i,q) & =\begin{cases}
    +1\quad\sigma_i^{(q)}=\uparrow \\
    -1\quad\sigma_i^{(q)}=\downarrow \\
    \end{cases}. \\
\end{align*}
We can note how the shifts in energy depend on the pairwise products of the spins' directions in the basis vector. 
In order to simplify our analytical computation, we consider regular networks where each node is connected to its first $k$ neighbors; arbitrary topologies introduced by the perturbation can be numerically computed from Eq.~\eqref{pert_general}. Starting with first, nearest-neighbors case, i.e. $\Delta_{ij}=\Delta_0$ if dist$(i,j)=1$ and vanishing otherwise, the energy shifts read:
\begin{align}    
    \langle\boldsymbol{\sigma}_q|\op{W}|\boldsymbol{\sigma}_q\rangle
    &= \frac{\Delta_0}{4}\sum_{i=1}^L \text{sgn}(i,q)\text{sgn}(i+1,q),
    \label{pert_shift_1} \\
L+1 &\equiv 1 \, \text{(PBC)}. \notag
\end{align}
Note that PBC stands for Periodic Boundary Conditions.

Instead, in the case of the first two neighbors having non-negligible interactions and within PBC:

\begin{align}    
    \langle\boldsymbol{\sigma}_q|\op{W}|\boldsymbol{\sigma}_q\rangle
    &= \frac{\Delta_1}{4} \sum_{i=1}^L \text{sgn}(i,q)\text{sgn}(i+1,q) \notag \\
    &\quad + \frac{\Delta_2}{4} \sum_{i=1}^L \text{sgn}(i,q)\text{sgn}(i+2,q),
    \label{pert_shift_2}
\end{align}

where $\Delta_1$ and $\Delta_2$ refer to the constant interaction coupling with the first and second neighbors, respectively.

The energy shift formula~\eqref{pert_shift_2} being similar in form to~\eqref{pert_shift_1}, we can hence turn to examine in detail the case of nearest-neighbor interaction for the perturbation. The shift reflects the fact that if there is a total number of $L$ couples (with PBC) and $b$ is the number of pairs with opposite spin directions in a given Pauli string, the sum of Eq.\eqref{pert_shift_1} corresponds to the number of consecutive pairs with spins pointing in the same direction after subtracting those with opposite spin, that amounts to $(L-b)-b=L-2b$.
By way of example, let us compare the resulting effect for the cases of one and two excitations. In the former case, there are only two out of the $L$ couples of spins that have opposite directions. Thus, the summation in Eq.\eqref{pert_shift_1} amounts to $L-4$ for each vector state of the restricted computational basis $\mathcal{P}_1$. This means that the shift is the same for both levels: importantly the perturbation has no overall effect on the local $z$-magnetization frequency. In the latter case, one can distinguish two possibilities for the Pauli strings, as depicted in table~\ref{table:exc_cases}. According to the considered vector, one can get different values for the summation in Eq.\eqref{pert_shift_1} and hence for the energy shift.

\begin{table}[!htp]
    \centering
    \caption{Multiple excitations in all-to-all connected XXZ network. Examples of energy shifts computation for one, two, and three initial excitations (rows), with 0(1) denoting a spin up(down). First column: case at study. Second column: corresponding example of Pauli string. Third and fourth columns reveal the summation and the shift computed from Eq.\eqref{pert_shift_1}, respectively.}
    \label{table:exc_cases}
    \begin{ruledtabular}
    \begin{tabular}{l c l c}
        Excitation(s) & Sample Pauli string & Summation & Energy shift \\
        \hline
        \textbf{1} & $\downarrow\downarrow\downarrow\uparrow\downarrow\downarrow\downarrow$ & $L-4$ & $\Delta_0(L-4)/4$ \\ [8pt]
        \textbf{2} $(b=2)$ & $\downarrow\downarrow\uparrow\uparrow\downarrow\downarrow\downarrow$ & $L-4$ & $\Delta_0(L-4)/4$ \\ [8pt]
        \textbf{2} $(b=4)$ & $\downarrow\downarrow\uparrow\downarrow\uparrow\downarrow\downarrow$ & $L-8$ & $\Delta_0(L-8)/4$ \\ [8pt]
        \textbf{3} $(b=2)$ & $\downarrow\downarrow\uparrow\uparrow\uparrow\downarrow\downarrow$ & $L-4$ & $\Delta_0(L-4)/4$ \\ [8pt]
        \textbf{3} $(b=4)$ & $\downarrow\uparrow\uparrow\downarrow\uparrow\downarrow\downarrow$ & $L-8$ & $\Delta_0(L-8)/4$ \\ [8pt]
        \textbf{3} $(b=6)$ & $\uparrow\downarrow\uparrow\downarrow\uparrow\downarrow\downarrow$ & $L-12$ & $\Delta_0(L-12)/4$ \\ [8pt]
    \end{tabular}
    \end{ruledtabular}
\end{table}

The present analysis can be generalized to an arbitrary number $w$ of excitations. Reverting back to the Pauli strings examples appearing in Table~\ref{table:exc_cases}, the largest shift is obtained when the configuration possesses the least amount of pairs with opposite spins, i.e. minimum number of domain walls: this implies $b=2$ and $\delta E_{\text{max}}=\Delta_0(L-4)/4$. Instead, the smallest shift is obtained when all the excitations are separated, or in other words $b=2w$ and $\delta E_{\text{min}}=\Delta_0(L-4w)/4$. Notice that we can now estimate the width of the band as the difference between the maximum and the minimum possible shifts, that is:
\begin{equation}
    \delta E_{\text{max}}-\delta E_{\text{min}}=\frac{\Delta_0}{4}(L-4-L+4w)=\Delta_0(w-1)\,.
    \label{maxshift}
\end{equation}
We thus see that after introducing a given number of excitations, the maximum shift interval in~\ref{maxshift} is proportional to the number of excitations in the quantum spin network. As a result, the quantum network develops an uncertainty in the encoding of the number of excitations, that scales linearly with the number of excitations itself. The physical understanding of this network property is that the anisotropic interaction between nearest-neighbor spins shifts the original energy levels, thus creating enlarged spectrum from degeneracy breaking with further and wider peaks.

This result can provide useful insights in the design of quantum technology applications embodying  multiplicative noise, i.e.  proportional to signal, thus leading to equal information per logarithmic interval of the variable~\cite{gardiner2004stochastic,nielsen2000quantum}. This can be relevant for the design of self-similar quantum networks or of non-classical states of matter for quantum metrology. Here, fluctuations due to phase diffusion, such as with collective spin states, cause the nature of noise to be multiplicative and represent the realistic limit to ideal quantum-sensitivity enhancement~\cite{demkowicz-dobrzanski2012metrology}.\\
Interestingly, this result can also be linked to the description of information processing properties of classical complex systems, performed with some quantum advantage by means of quantum network models. One such example concerns the numerosity perception in neuroscience, that is the ability of humans and mammals of counting the number of items in a given space without cognitively counting, performing with an about 15\% relative uncertainty in a quite large range of items number, known as Weber's law~\cite{dehaene2003weberlaw}. This phenomenology is experimentally observed under very general conditions, no matter the perception channel (i.e. visual or auditory), nor the way the items are presented~\cite{burr2008visual}. Importantly, neuroscience experiments indicate that the perception of number is interconnected to the perception of time and space intervals~\cite{Morrone:2007}, challenging neuroscience understanding. In~\cite{YagoMalo2023,yagomalo2024review}, the mapping of the problem into an out-of-equilibrium XXZ network with all-to-all connectivity has been shown to account for the Weber's law in effective and efficient manner, whereas conventional artificial neural networks poorly perform~\cite{nasr2019deepneural}. There, an ideal-observer analysis of the power spectra had shown that the uncertainty with which the quantum network encodes the number of frequency peaks - and therefore the number of excitations - linearly scales with the number itself. The analytical understanding that we have built up in the present work  provides neuroscientists with useful insights and can be promising to link numerosity perception with time and space-intervals perception.

With this physical understanding at hand, we can now turn our attention to the transport properties of more general quantum networks, where we release either the constraint of constant coupling intensity of the links in the regular network, or the regular architecture itself by considering complex quantum networks. This is the subject of the next Section~\ref{sec:trans_gen_net}, where we explore the extent to which the space and time memory-preserving property of the regular XXZ network excitations can be engineered in more general networks; We will focus, in particular, to the case of one initial excitation or, more generally, in the $m=-L/2+1$ total $z$-magnetization sector.

\section{Transport properties in generalized quantum networks}
\label{sec:trans_gen_net}

So far we built a theoretical understanding about the out-of-equilibrium dynamical behavior of the closed quantum spin network governed by the XXZ Hamiltonian, for a network with all-to-all connectivity architecture of equally weighted links. For this case, we could resort to semi-analytical means. In this Section, we start from the understanding developed in Section~\ref{sec:symm_and_dyn} to numerically investigate the role played by (i) the interaction graph in a complex quantum system architecture, and (ii) the range of the interactions according to a power law when all sites are connected between them. For simplicity, we stick to the case of one single excitation, having developed in Section~\ref{sec:symm_and_dyn} an intuition for multiple excitations: we will come back to this point in the concluding Section~\ref{sec:conclusion}. As an example for complex quantum network, we choose the small-world architecture: besides being \textit{per se} relevant for diverse problems, we are here interested to its property of well interpolating between the cases of regular and random graphs after tuning only two parameters.

We analyze the underlying physics by specializing our analysis to a localization quantifier called Inverse Participation Ratio (IPR), see Appendix~\ref{appdx:ipr_def}. The analysis that follows will follows on the XX part of Hamiltonian~\eqref{hamiltonian1}, i.e. all coefficients $\Delta_{ij}$ will be set to 0.

\subsection{Small-world networks}
A number of real-life networks relevant to diverse phenomena, ranging from biology to social sciences, are characterized by a high level of clustering and, at the same time, by a low average distance between the vertices: this is the so-called “small-world phenomenon” that illustrates e.g. social networks and the World Wide Web \cite{watts1998smallworld}.

Such a network can be built from a regular-lattice graph where each node is linked to its first $k$ neighbors; going clockwise through all nodes, for all $k$ links situated on the left of it we rewire it with probability $\beta$ to a different one.
For the types of network we are considering, $k$ can vary from 1 to $\lfloor L/2 \rfloor$; related to it, the parameter $\bar{g}=2k$, ranging from 2 to $L-1$ by steps of 2, drives the level of network complexity, as it denotes the number of initial connections for each site and also the average degree defined below, due to the way these networks are constructed.
The parameter $\beta$ governs the level of graph's randomness, bridging between the regular-lattice case with $\beta=0$ and the fully random graph with $\beta=1$. The all-to-all connected graph is obtained in the limit $\bar{g}=L-1$, while the nearest-neighbor case has $\bar{g}=2$. The construction mechanism is illustrated in Fig.~\ref{fig:sw-magz}$(a)$. Throughout the rest of the article we treat the case of odd $L$ for ease of notation without loss of generality.

We complete this introductory Section by recalling the main qualifiers and quantifiers for the networks architectures considered in complex-network theory, that will be used in the following. For details, we refer to the Appendix~\ref{appdx:avg_dist_clust}. One first connectivity quantifier is the node degree, denoted by the word ``deg", that is the number of edges linking one node to others. The coupling range is quantified by the topological distance dist$_{ij}$, counting the number of links connecting two sites in their shortest path: a global distance quantifier $\mathcal{D}$ can then be usefully introduced, as the average of the topological distance over all the pairs of nodes. Finally, the tendency to aggregate is quantified by the global clustering coefficient $\mathcal{C}$, that is the average proportion of total triangles (meaning closed loops of three vertices) within a given neighborhood, as related to the maximum number of allowed connections according to the local degree. All-to-all connected graphs are the limiting case of low topological distance, i.e. tending to 1, and high clustering coefficient, i.e. tending to 1; in the opposite limiting case of nearest-neighbor networks both local and global clustering are 0, since no triangles between nodes can be observed, while average distance linearly scales as the number of sites $L$.

\subsection{Memory effects in small-world networks}
\label{sec:msm}

\begin{figure*}[!tb] 
    \centering
    \includegraphics[width=\textwidth]{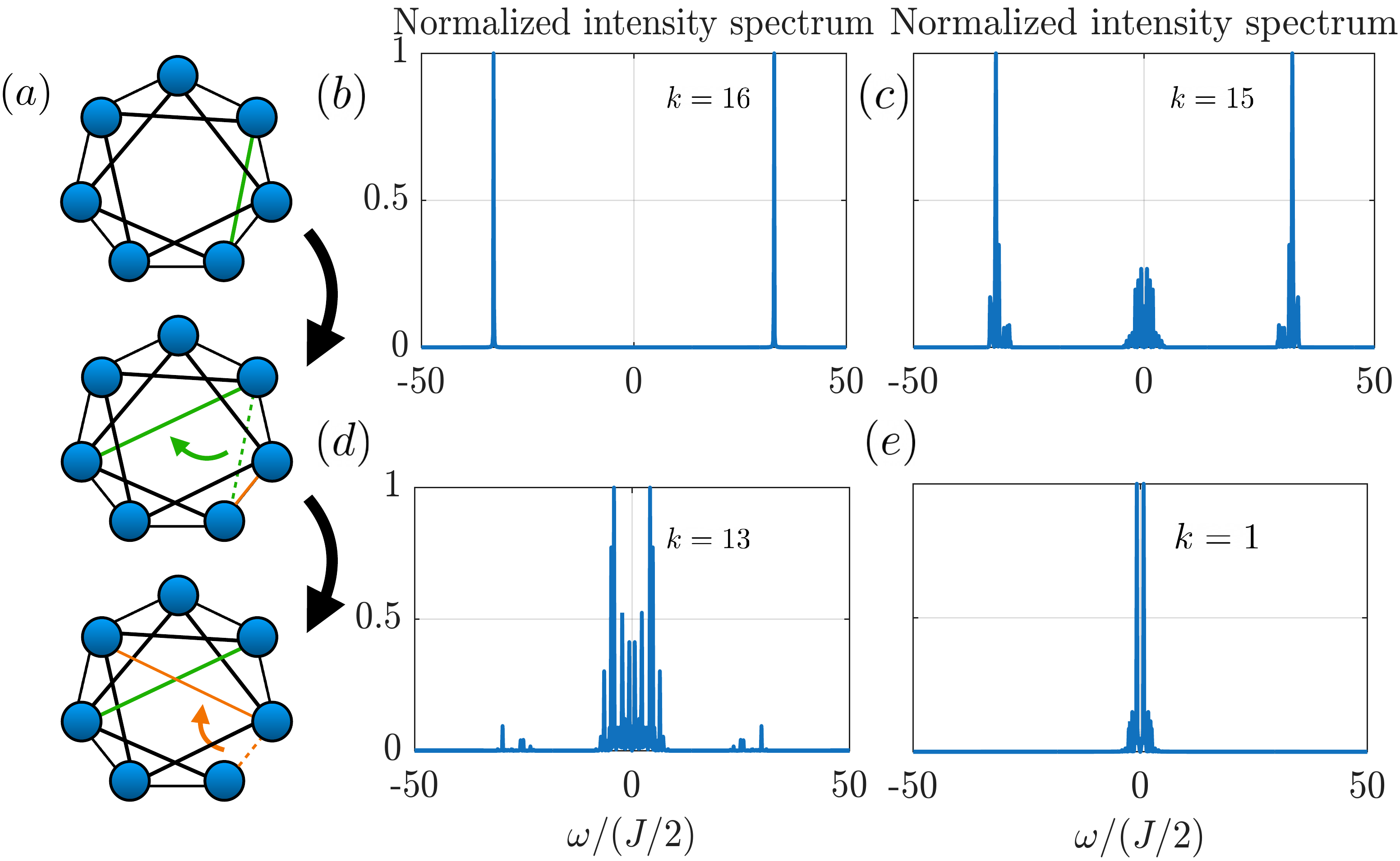}
    \caption{  
    Construction of small-world graphs and their memorization properties. $(a)$ Small-world network construction (see text). Top to bottom: a regular graph with $L=7$ sites and $k=2$ nearest-neighbor connections, and different rewired links (green and orange) with respect to the original graph, according to the value of the probability $\beta$.
    $(b)-(e)$ Examples of intensity spectra for average $z$-magnetization at the initial site for one quenched excitation along the $z$-axis in four different networks with $L=33$ sites. $(b)$ All-to-all network, with delta-like peaks at frequencies $\omega^*=\pm JL/2$=33, see Eq.\eqref{freq_spliten}. $(c)$ Small-world, highly-connected graph with $(k,\beta)$=(15,0.5), possessing additional low-frequency components. $(d)$ Small-world, highly-connected graph with $(k,\beta)$=(13,0.5), having small high-frequency contributions. $(e)$ Sparsely connected small-world with $(k,\beta)$=(1,0.2), having no high-frequency peaks. The data has been detrended before the Fourier transform by removing the zero-frequency component.}
    \label{fig:sw-magz}
\end{figure*}

Here we analyze the memory effect discussed in Section~\ref{sec:xx-ham} in small-world networks, starting with a single excitation, i.e. a single spin oriented along $z$ while all the other spins are along the opposite direction. We use the same technique of Section~\ref{sec:xx-ham} in which we take the Fourier transform of $S^z_{\text{initial}}(t)$. In presence of a single excitation only one high-energy resonance (whose energy scale with the system size) is present in the dynamics for all-to-all interaction, allowing us to count the number of excitations. Here we analyze how the Fourier spectrum changes as we decrease the connectivity. Interestingly, even if the number resonances increases as we decrease the connectivity, the high-energy resonances persists even for lower connectivity, see Fig.~\ref{fig:sw-magz}$(c)$. This result could allow to extend the notion of memory effect to generic small-word network. Indeed, it shows that signatures of the memory-preserving effect survive even for less connected network. Clearly, it would be interesting to study if the memory-preserving effect persists in small-world networks also in the many excitation case. This will be the topic of a future work.

\subsection{Inverse Participation Ratio}
\label{sec:ipr}

The transport properties of one excitation in such networks can be characterized by means of a delocalization quantifier, the Inverse Participation Ratio (IPR)
\begin{equation*}
    \text{IPR}(t)=\sum_s \langle s|\op{\varrho}(t)|s\rangle^2\,, 
\end{equation*}
defined as the sum of the squared trace elements of the density matrix $\op{\varrho}$ over single-excitation localized states {$\{|s\rangle\}=\{|\downarrow\downarrow\dots\downarrow\uparrow\downarrow\dots\downarrow\rangle\}$, where the $\uparrow$ in the previous Pauli string is a spin up (along $z$) located at site $s$; this basis has dimension $L$ and constitutes one of the subspaces of the total $z$-magnetization operator $\op{M}_z$ with which $\op{H}$ can be block-diagonalized. While we refer to Appendix~\ref{appdx:ipr_def} for a complete definition, we here highlight that by definition the IPR values range from $1/L$ for the complete delocalized state, i.e. infinite-temperature case, to $1$ for the perfect localization on a single eigenstate. If needed, generalizations of this quantifiers to the case of multiple excitations can be introduced, as in~\cite{Civolani2024}.

Before proceeding, it is useful to briefly discuss the form of the IPR function for a single excitation in the all-to-all case, as in Section~\ref{sec:symm_and_dyn}: following the theoretical discussion in Section~\ref{sec:xx-ham}, one can easily derive the analytical function as the sum of a constant term and two cosine functions, oscillating with frequencies $\omega^*$ and $2\omega^*$, as $\omega^*$ is the same frequency of the local $z$-magnetization signals that is the energy gap between the levels $|l=L/2,m=-l/2+1\rangle$ and $|l=L/2-1,m=-L/2+1\rangle$, see Eq.\eqref{freq_spliten}. Therefore, in the following we will also show the IPR average for the all-to-all case over a temporal window $2\pi/\omega^*$. The way delocalization happens in small-world networks is strongly dependent on the initial conditions, which for our case translates to the site of the first excitation and its degree; this is the focus of Section~\ref{sec:init-cond}.

\subsection{Dependence on initial conditions}
\label{sec:init-cond}

We start by seeking an  understanding of the importance of initial conditions, as we have done in Section~\ref{sec:symm_and_dyn}. Since the analytical machinery that we established for regular networks seems to not apply for irregular graphs, we seek for alternatives. It turns out that this is possible at least for  the short-time IPR behavior, by means of the so-called cascade model introduced in \cite{borgonovi2019ipr}: in essence, this monitors the initial state $|x(0)\rangle=|s\rangle$ coupling with other basis elements at different moments $|x(t)\rangle$; here, $|s\rangle$ denotes the initial site of the excitation.

The underlying idea is as follows, see Fig.~\ref{ipr-sw-cascade}$(a)$ for the concept. At $t=0$, only the subset of state vectors $\mathcal{F}_0$ containing $|s\rangle$ participates in the temporal evolution. Immediately after, to first-order approximation the temporal evolution operator connects the initial state to the basis vectors $|f_i\rangle$ for which the matrix elements $\langle f_i|\op{H}|s\rangle$ are non-vanishing, thus extending the subset of involved states to $\mathcal{F}_1$. More precisely, the subset of states grows as:
\begin{eqnarray}
    \mathcal{F}_0 && =\{|s\rangle\} \xrightarrow{\exp{(\mathrm{i}\op{H}t)}\simeq (\hat{\mathbb{I}}-\mathrm{i}\op{H}t)}\mathcal{F}_1 \nonumber \\
    && =\{|s\rangle\}\cup\{|f_i\rangle\, ,\,\text{$f_i$ first neighbors of $s$}\}\,.
\end{eqnarray}

\begin{figure*}[] 
    \centering
    \includegraphics[width=\textwidth]{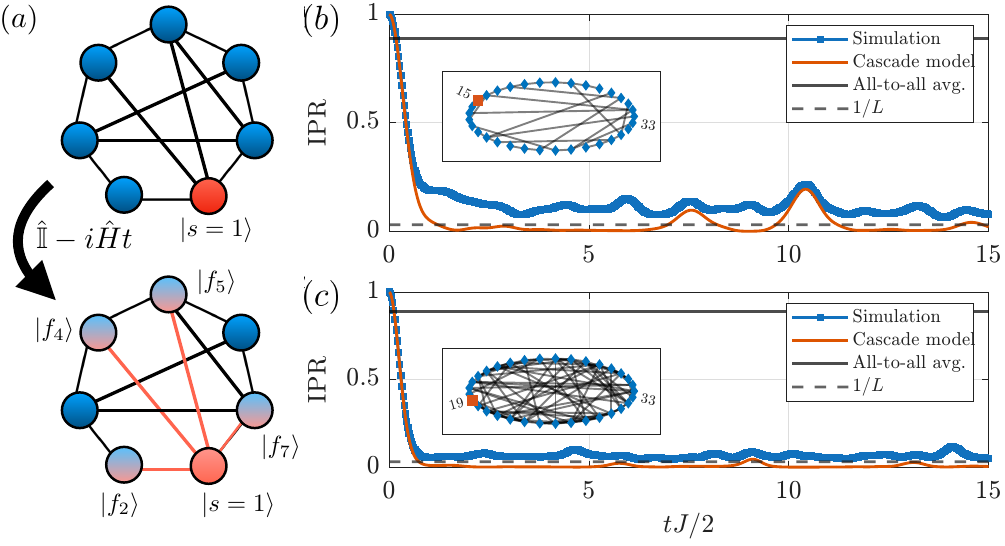}
    \caption{Understanding transport in small-world graphs at early times. $(a)$ Concept of the cascade model (see text). Top: an initial excitation (red circle) in a generic network (black edges) is located at site 1, as denoted by the state $|s=1\rangle=|\uparrow\downarrow\dots\downarrow\rangle$ and the red circle opposed to the blue ones. Bottom: after a short temporal evolution (black arrow) the initial excitation has spread (gradient-colored circles) across site 1's first neighbors (orange edges), such that the state is a superposition of $|s=1\rangle$ and states $|f_2\rangle=|\downarrow\uparrow\downarrow\dots\downarrow\rangle$, $|f_4\rangle$, $|f_5\rangle$ and $|f_7\rangle$. $(b)-(c)$ Comparison of the IPR's time evolution as computed from the cascade model (orange curves) and from a numerical simulation (blue curves) for the following small-world networks: $(b)$ $k=1$, initial site with degree 2; $(c)$ $k=3$, initial site with degree 4.}
    \label{ipr-sw-cascade}
\end{figure*}

In fact, one can set and resolve rate equations for the probabilities of finding the state evolution $|x(t)\rangle$ to a particular basis element $|f_i\rangle$:

\begin{align*}
    P_{s}(t) & =|\langle s|x(t)\rangle|^2\,, \\
    P_{f_i}(t) & =|\langle f_i|x(t)\rangle|^2\,.
\end{align*}

Borgonovi et al. have found a transparent expression for the evolution of IPR at short times, in the approximation where any back-flow to the initial site can be neglected~\cite{borgonovi2019ipr}. This is:
\begin{equation}
    \text{IPR}(t)\approx P_s^2+\frac{\sum_{f_i} P_{f_i}^2}{\mathcal{N}_1}\,,
    \label{ipr_short}
\end{equation}
where the sum is performed over the first-neighbor sites of the initial one, and $\mathcal{N}_1$ is the number of such nodes plus one (referring to vertex $s$). Let us better understand Eq.\eqref{ipr_short}: as the survival probability $P_s$ decreases because of the excitation spreading, the initial site couples with its first neighbors in the interaction topology. In other words, the IPR evolution is strictly dependent on the graph upon which it lives, through its adjacency matrix $A_{ij}$ describing the existing connections and being proportional to the XX Hamiltonian itself, when written in the restricted basis of computational vectors having one excitation. Notice that $A_{ij}=1$ if sites $i$ and $j$ are linked, else $A_{ij}=0$. Under a different perspective, the degree of the initial node (i.e. the number of first neighbors in the graph) governs the transport onset at short times. 

This procedure can easily be generalized to later times. Indeed, as we go further, the powers of $\op{H}$ participate to the initial state evolution. Since $\op{H}$ is proportional to the adjacency matrix, elevating the latter to a certain power $p$ tells us the topological distance between two sites if it is less than $p$, i.e. it refers to the subset of states contributing to the time propagation.

A summary of the analytical results is presented in Fig.~\ref{ipr-sw-cascade}. In Fig.~\ref{ipr-sw-cascade}$(b)-(c)$, we compare the simulated (blue curves) IPR time behavior with the predictions of the cascade model (orange curves) for two complex network graphs that are sparsely $(b)$ and highly $(c)$ connected, respectively. For earlier times we find a good agreement between the IPR simulations and the predictions provided by the cascade model. In fact, this is able to predict the sharp deviation from the all-to-all case, heralding the breaking of the spectral degeneracy and the halt of cooperative shielding. The later times behavior from the cascade model misses the bumps that are likely due to the backflow processes, that are neglected in the model approximations.

Another illuminating way of looking at the impact of initial conditions on excitation spreading regards increasing connectivities close to the limit of all-to-all connected networks. The results are illustrated in Fig.~\ref{ipr-sw-high-conn}$(a-b)$, where we consider a network with a given $\bar{g}=L-3$ parameter, closest to all-to-all case but not-fully connected, and impose a non-zero rewiring probability providing the sites with different degrees, i.e. different number of links with other elements. Then, we compute the IPR by picking different initial sites according to their degree. We do this for both long, Fig.~\ref{ipr-sw-high-conn} $(a)$, and short, Fig.~\ref{ipr-sw-high-conn} $(b)$ temporal windows, and we observe that the IPR time-evolution behavior for the all-to-all case soon breaks into the general small-world system behavior. Here, the degeneracy lifting due to the symmetry reduction with respect to the all-to-all case makes the IPR decreasing over time. The stationary value of the IPR critically depends on the degree of connectivity of the initial site.

We quantify the departure of small-world behavior from the all-to-all connectivity case in Fig.~\ref{ipr-sw-high-conn}$(c)$, where we introduce an indicator for the difference the distance:
\begin{equation}
    \Delta_{\text{IPR}}(t)=|\text{IPR}_{\text{a.a.}}(t)-\text{IPR}_{\text{s.w.}}(t)|\,.
\label{eq:IPRdistance}
\end{equation}
Different dots colors for $\Delta_{\text{IPR}}$ refer to values computed at different time instants $tJ/2$ in units of $2\pi/\omega^*$, $\omega^*$ being determined as in Eq.\eqref{freq_spliten}; for these values, IPR$_{\text{a.a.}}$=1. Each point represents the average computed over initial sites having the same degree in the same  network as in Fig.~\ref{ipr-sw-high-conn}$(a)$. As expected, we observe an increasing difference while the network's degree departs from $L-1$ (degree/$(L-1)<1$) corresponding to all-to-all connectivity, the difference being enhanced at each degree value with increasing time.

\begin{figure*}[] 
    \centering
    \includegraphics[width=\textwidth]{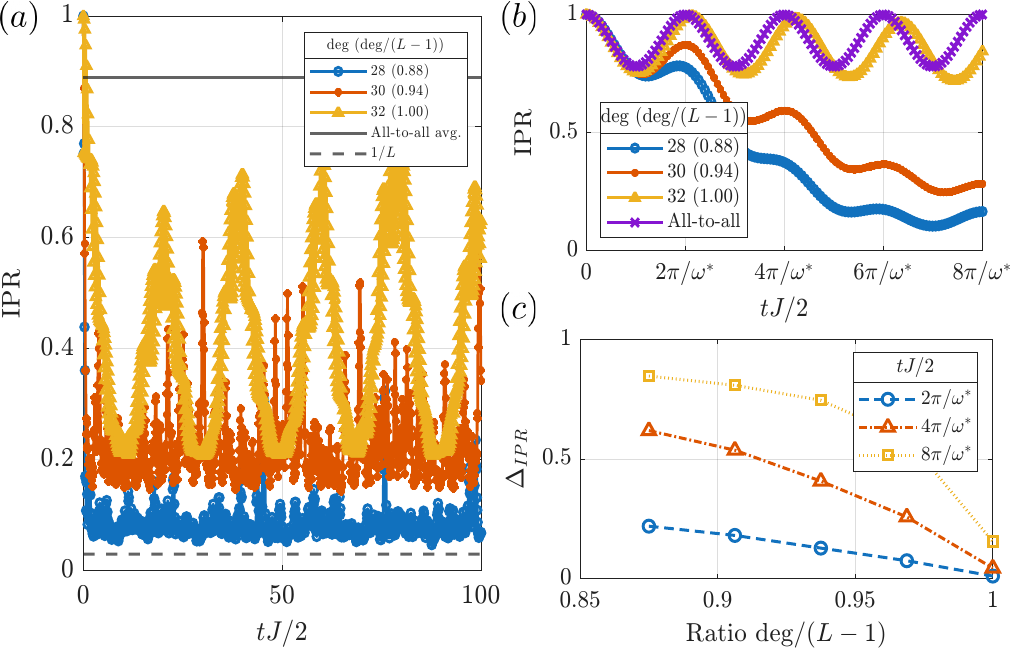}
    \caption{Understanding transport in highly-connected networks. $(a)$ IPR behaviors for long periods of time for varying initial sites, denoted by their degree, in a fixed network with $L=33$, $\bar{g}=L-3$, $\beta=0.1$. $(b)$ Same as $(a)$ but for short times; here $\omega^*=JL/2=33$ dictates the oscillation periods of the all-to-all connected IPR. $(c)$ IPR distance \eqref{eq:IPRdistance} between all-to-all and small-world networks graphs as a function of the initial site's degree related to $L-1$ at increasing times as in the legend.}
    \label{ipr-sw-high-conn}
\end{figure*}

\subsection{Infinite-time localization depending on graph properties}
We thus turn on investigating the IPR for graphs highly connected on a global scale, after tuning the tendency to clusterize and the rewiring probability. In so doing, we aim to study the spreading of information and its dependence on the system  parameters.  Therefore, we now revert back to simulations performed over longer-enough temporal windows, so that information or energy has spread throughout the network, to evaluate the participation of the basis vectors to the initial state’s temporal evolution (given by its IPR). To explore the whole physics landscape, we now simulate a set of 20 graphs with size $L=33$, for each pair of generative parameters $(k,\beta)$ with $k=1,2,\dots,15$ (maximum possible $k=16$, equivalent to the all-to-all situation) and $\beta=0.1,0.2,\dots,1$. For each case study, we take the long time average value of the IPR, computed for an initial site chosen at random, in a temporal window chosen such that the excitation could have the time to travel across the whole graph: specifically, we simulate the temporal evolution for the period $tJ/2\in[0, 100]$ and compute the average for times in the interval $tJ/2\in[90, 100]$.

We summarize the results in Fig.~\ref{sw_ipr_merit}. The boxplot, Fig.~\ref{sw_ipr_merit}$(a)$, displays the average IPR as a function of the number of initial neighbors $\bar{g}$, for different values of the rewiring probability $\beta$. We see that localization, i.e. higher IPR values, is favored for extremal values of $k$ (1 and 15). The lower extreme  corresponds to the obvious case of low connectivity, specifically similar to a nearest-neighbor connection. The latter, instead, corresponds to the case of highly-connected graphs with significant interference effects. Delocalization manifests instead at intermediate connectivity conditions. The rewiring probability $\beta$ does not significantly influence the average value, essentially dominated by the number of neighbors. We associate variations in this behavior with the limited number of sampled graphs. 

We now represent in Fig.~\ref{sw_ipr_merit}$(b)$ the same IPR results in a scatter plot where the graphs are characterized by their average distance and average clustering instead, see Appendix~\ref{appdx:avg_dist_clust}. We see that relatively localized long-time states are possible away from the limit of all-to-all connected graphs, that is with average clustering and average distance equal to 1. Quite interestingly, non-trivial steady states emerge not only in the highly connected cases but also in sparsely connected graphs with high-average distance (lower-right part of the plot), due to the slow propagation of excitation and small interference. This behavior heralds the potential for a memory-preserving property of the initial quench even away from the all-to-all case, should a certain decoding protocol be provided. In contrast, with $k$ values  comparable to the all-to-all case with $L(L-1)/2$ (upper-left part), signatures of cooperative shielding appear, typical of complete connection. The difference between the finite IPR values (green to light-blue symbols) shown in the top-left and in the bottom-right part of Fig.~\ref{sw_ipr_merit}$(b)$ is better envisioned by inspecting Fig.~\ref{sw_ipr_merit}$(c)$. Here we show a typical IPR vs. time $tJ/2$ behavior (blue line) for parameter values in the bottom-right part of $(b)$ for $k=1$. The IPR signal oscillates around an average (dot-dashed horizontal line) that is clearly above the threshold $1/L$ (dashed horizontal line) of complete delocalization, though the IPR has a lower value with respect to the IPR in a corresponding all-to-all case (yellow). The cascade model prediction (red) only provides a reasonable fit at all times. The differences are due to the fact that the cascade model neglects all backflow processes. Thus, we might infer that the emerging localization in this region is enhanced by quantum interference effects among different paths, in a mechanism similar to Anderson localization~\cite{alet2018_mbl}. Finally  we note that the question of whether cooperative shielding survives in more general networks deserves further study which requires a scaling analysis as a function of the system size. This topic will be analyzed in a future work.

\begin{figure*}[]
    \centering \includegraphics[width=\linewidth]{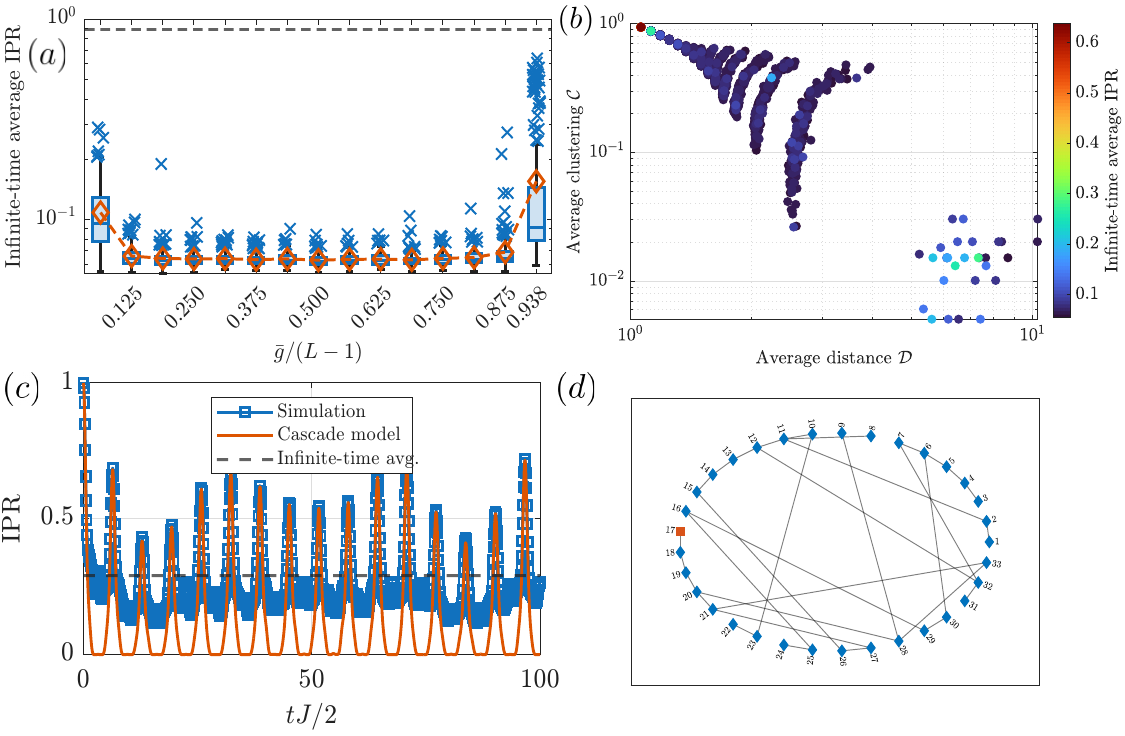}
    \caption{Effect of network topology on the long-time average IPR. For each pair $(k,\beta)$ of initial conditions in the graph generation, the computed average IPR in the time window $tJ/2\in[90, 100]$, to ensure that the excitation could travel across the whole graph. Each pair is tested 20 times and each time the initial site is chosen at random.
    $(a)$ Long-time averaged IPR vs. the number of initial neighbors $\bar{g}$ normalized to $L-1$. The horizontal dashed line on top is the average IPR for the all-to-all case. For each $\bar{g}/(L-1)$ value, boxes represent the data's interquartile range, i.e. all points whose mean IPR belong between the 25\textsuperscript{th} and 75\textsuperscript{th} quartile, and have a horizontal line marking the median; whisker lines extend up to 1.5 times the interquartile range outside of boxes; crosses depict the outliers, or points that are not included in neither boxes or whiskers; finally, the red diamonds display the average IPR sampling. Localization is seen to occur at extremal values of $k$, i.e. nearest-neighbor and all-to-all type of connectivity. Parameter values: $k=1,\dots,15$, $\beta=0.1,0.2,\dots,1$, $L=33$. $(b)$ Long-time averaged IPR in the form of a scatter plot in the parameter space of average clustering $\mathcal{C}$ and average distance $\mathcal{D}$ (data having $\mathcal{C}=0$ is not shown). The general trend going from lower-right to upper-left denote increasing values of $k$. Notice that, away from all-to-all connectivity with $\mathcal{C}=\mathcal{D}=1$, some degree of localization is still possible for more regular networks, i.e. sparse graphs with large $\mathcal{D}$ and intermediate to very low values of clustering $\mathcal{C}$. We see this in panel $(c)$ showing a typical graph of IPR vs. dimensionless time $tJ/2$ (blue line) for parameter values $k=1, \beta=0.5, \mathcal{D}=7.0, \mathcal{C}=0$ in the bottom-right part of $(b)$, corresponding to the network in $(d)$ initialized at site $17$, as depicted. The IPR displays irregular oscillations around an average (dashed line) that is well above the threshold $1/L$ of complete delocalization. For comparison, the IPR signal is shown with the respective cascade model (red).}
    \label{sw_ipr_merit}
\end{figure*}

\subsection{Power-law-dependent hopping}
A different way to break space regularity in networks is to consider exchange coupling that vary with physical distance, specifically characterized by a power law, i.e. $J_{ij}\propto\text{dist}(i,j)^{-\alpha}$ in Eq.\eqref{hamiltonian1} which in the case of PBC is equivalent to min$(|i-j|,L-|i-j|)$. This test case constitutes a good bench to showcase cooperative shielding, introduced in Section~\ref{sec:intro}, when considering different initial conditions: in fact, on the one hand increasing values of the power law coefficient cause shorter ranges for the interactions (as was for diverse combinations of $k$ and $\beta$ for small-worlds), and on the other hand each site is equivalent to the others (unlike the previous varying network case). Here we analyze the memory effects in such models and the dynamics focusing on the cooperative shielding effects. 

\subsubsection{Memory effects}
Here we analyze the memory-preserving effect discussed in Section~\ref{sec:xx-ham} in networks with power-law interactions, starting with a single excitation and taking the Fourier transform of $S^z_{\text{initial}}(t)$, see Fig.~\ref{fig:powerlaw-spectra}$(a)$.  As we increase $\alpha$ the high-energy frequencies are reduced but they still persist.  Also in this case, this result could allow us to extend the notion of memory-preserving effect to generic range of hopping. 

\subsubsection{Dynamics/shielding}
Here we study the dynamics for different initial conditions, considering also  superpositions of single-excitation states. This is illustrated in Fig.~\ref{fig:ipr-powerlaw}$(a)$, where we compare the IPR evolution at early and late times for positive values of $\alpha$ (meaning decaying coupling with distance) and two possible initial conditions: the first one is the excitation localization in one site (s.s.), such as $|\uparrow\downarrow\dots\downarrow\rangle$, the second is an antisymmetric quantum superposition (a.q.s.) of the kind ($|\uparrow\downarrow\dots\downarrow\rangle$-$|\downarrow\uparrow\dots\downarrow\rangle$)/$\sqrt{2}$; note that in the latter case the initial superposition is spread across sites which are nearest neighbors, with distance min$(|i-j|,L-|i-j|)$=1. The vertical axis represents IPR normalized with respect to its initial value, which is 1 for single site and 0.5 for the superposition considered. We can clearly see that the latter initial condition, a.q.s., allows for longer (relative) localization, meaning that a significant decay of the function happens at much longer times for antisymmetric superposition rather than pinpointed localization states. This is even more evident when we try to compute this time span: to do so, we look for the stopping time $\tau_{1/2}$ for which normalized IPR becomes 1/2 for both initial states, and then we compute the difference:
\begin{equation}
    \Delta\tau_{1/2}=\tau_{1/2}^{a.q.s.}-\tau_{1/2}^{s.s.}\,.
\end{equation}
The results are displayed in Fig.~\ref{fig:ipr-powerlaw}$(b)$, where we obtain $\Delta\tau_{1/2}$ for different values of $\alpha$ and different scales or number of sites $L$. We first notice that the stopping time difference is a decreasing function of $\alpha$ and reaches the lower bound corresponding to the nearest-neighbor network case already for $\alpha\gtrsim 5$.

\begin{figure*}[] 
    \centering
    \includegraphics[width=\textwidth]{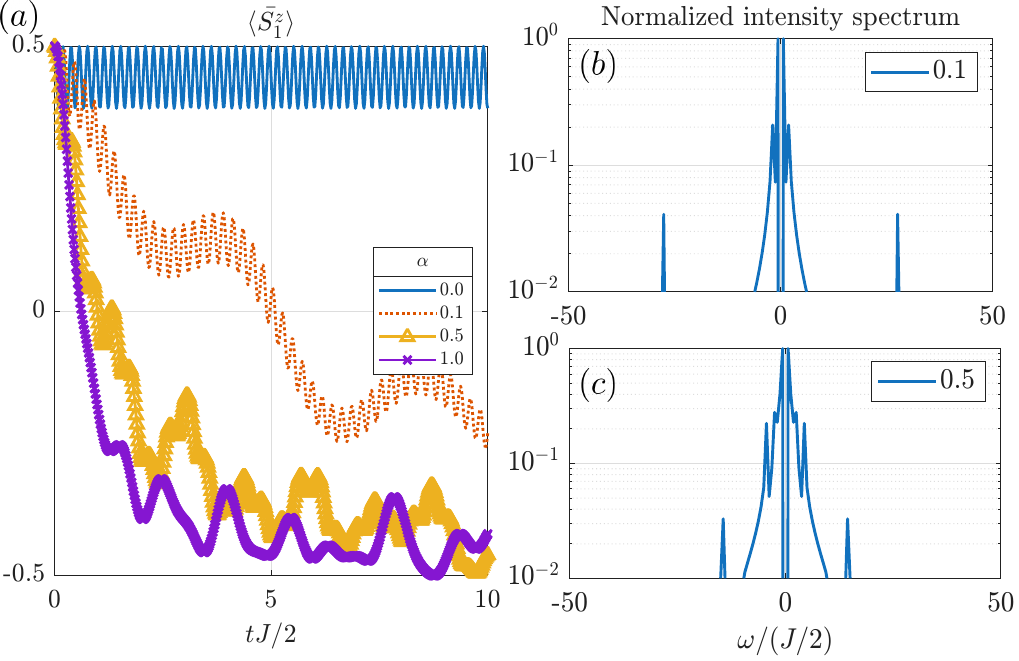}
    \caption{Memorization in all-to-all connected graphs with varying interaction range. $(a)$ Average local $z$-magnetization at the initial site for different power-law exponents $\alpha$ regulating the range of the couplings. The network possesses 33 sites and $J/2$=1. $(b)-(c)$ Corresponding power spectra for the signals depicted in Panel $(a)$ with $\alpha$=0.1,0.5; for comparison, the case $\alpha=0$ is shown in Fig.~\ref{fig:sw-magz}$(b)$. The data has been detrended before the Fourier transform by removing the zero-frequency component.}
    \label{fig:powerlaw-spectra}
\end{figure*}

The results shown in this Section are consistent with what was claimed in \cite{celardo2016_shielding} where it was shown that single excitation transport in presence of long-range hopping is highly dependent on the initial conditions. The results presented here confirm that this is a general effect of long-range hopping models. Indeed for short-range hopping, $\alpha \gg 1$ no dependence on the initial condition is shown, see Fig.~\ref{fig:ipr-powerlaw}$(a)$. Note that the antisymmetric initial state has a large overlap with the shielded subspace~\cite{celardo2016_shielding}, even if it does not lie completely in the shielding subspace and this could explain the lack of cooperativity (dependence on the size of the system) shown in Fig.~\ref{fig:ipr-powerlaw}$(b)$. The data has been detrended before the Fourier transform by removing the zero-frequency component.}

\begin{figure*}[] 
    \centering
    \includegraphics[width=\textwidth]{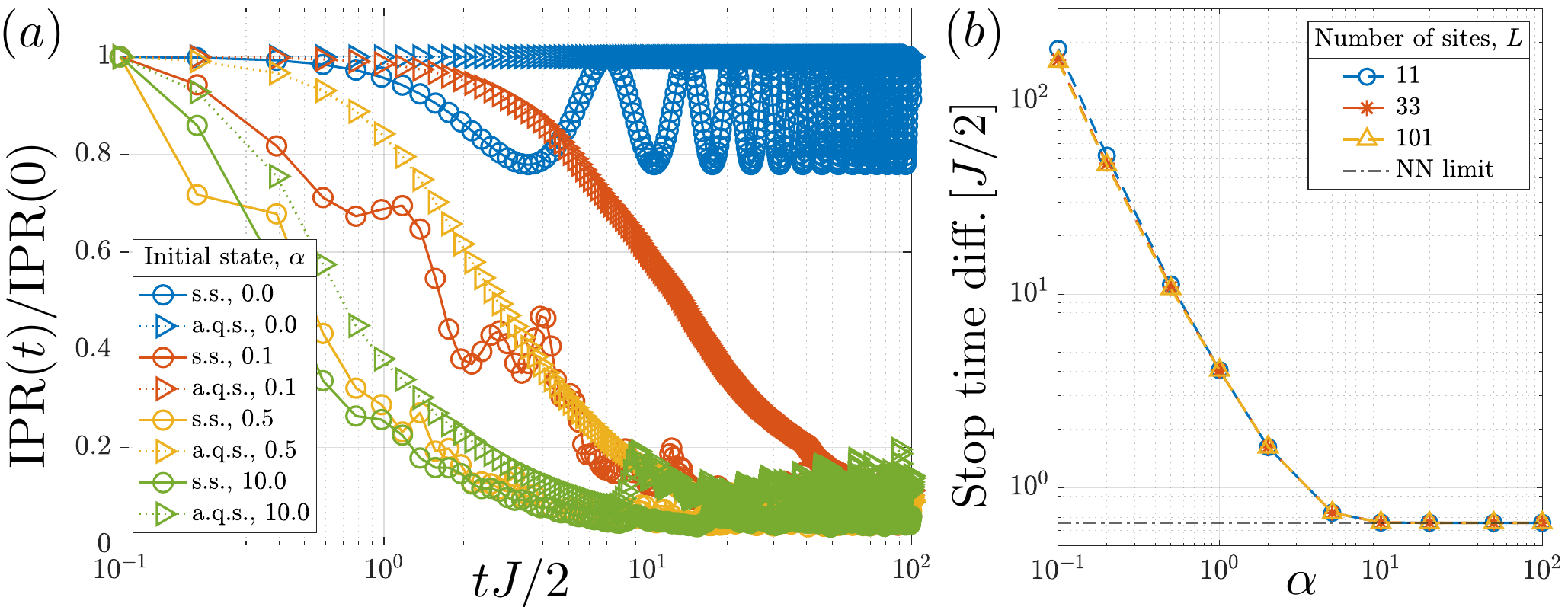}
    \caption{Dependence of transport on initial conditions for power-law decaying couplings. $(a)$ IPR time evolution for a fully-connected network with $L=33$ nodes and links intensity that changes with physical distance according to a power law with coefficient $\alpha$ as in the legend, $J_{ij}\propto \text{dist}(i,j)^{-\alpha}$, where dist$(i,j)=$min$(|i-j|,L-|i-j|)$ for the PBC considered in all these cases. Curves with continue stroke and circular markers denote the initial state having a well localized excitation (s.s.) $|\uparrow\downarrow\dots\downarrow\rangle$, while dashed lines with triangular markers symbolize the initial state with antisymmetric quantum superposition (a.q.s.) ($|\uparrow\downarrow\dots\downarrow\rangle$-$|\downarrow\uparrow\dots\downarrow\rangle$)/$\sqrt{2}$. The IPR function is normalized to the initial value, i.e. 1 for the former state and 0.5 for the latter. $(b)$ Difference in stopping time between a.q.s. and s.s. as function of $\alpha$. Dashed curves with different colors and markers represent results obtained for different system sizes $L$ and power-law exponents in the range [0.1, 100], and the lower bound indicate the limit of nearest-neighbor network, independent of scale.}
    \label{fig:ipr-powerlaw}
\end{figure*}

\section{Discussion and conclusions}\label{sec:conclusion}
This work is motivated by the relevance of complex quantum networks in the modeling of biological systems and as a powerful tool to extract relevant information for many-body systems. We have therefore addressed the transport properties of quantum networks as governed by different parameters such as the connectivity graph, the topology and  the range of the couplings connecting the nodes. These parameters can be tuned in current quantum technology platforms~\cite{yagomalo2024review}. We have focused on networks described by the XXZ Hamiltonian, that is a paradigmatic model with a variety of applications~\cite{franchini2017integrable}. 
To gain insights on this complex matter, we have adopted a methodology that progressively develops understanding from the simpler to harder cases: we start from the all-to-all connectivity with constant coupling among the nodes, then introducing power-law interactions in the regular network, and eventually passing to irregular, complex networks. The methodology employs combined analytical and numerical methods, revolving around the cooperative shielding concept and quantum cascade model~\cite{santos2016cooperative} accompanied by Exact Diagonalization. 
The main focus of this manuscript is on the effect of long-range interactions and connectivity on the dynamics of the network. We show that long-range coupling in highly connected networks imposes strong constraints on the dynamics which manifest themselves in the memory-preserving and cooperative shielding effect~\cite{santos2016cooperative,celardo2016_shielding}. 

Motivated by the findings in~\cite{YagoMalo2023}, we have analyzed the robust preservation of the memory of the number of initial excitations over long times. This peculiar phenomenon essentially results into a spatiotopic representation of the number of frequencies in the power spectrum of the network, that are as many as the number of excitations initially injected.  Even more interestingly, the uncertainty in the number of excitations that the quantum network can map, linearly increases with the number itself. This behavior is shown in~\cite{YagoMalo2023} to elegantly account for the so-called Weber's law for numerosity perception in neuroscience problems, after mapping the essential functions of the classical neuronal network into an open quantum network (resulting into the XXZ model), and in contrast with the poor performance of conventional biologically inspired models like artificial neural networks. In fact, this memory preserving property can be of much interest for concrete implementations with current quantum technologies. 

Three main results emerge from our analysis. First, the memory preserving properties of the all-to-all connected network have been traced back to the symmetry structure of the Hamiltonian. Both the one-to-one correspondence of the number of excitations with the number of frequencies in the power spectrum and the number uncertainty have been analytically explained, the former as a consequence of the Wigner-Eckart theorem to get the allowed quantum numbers from the usual angular momentum composition rules. 
The memory preserving effect has been analyzed also in more irregular networks, such as small world network and power law connected network. We have shown that signatures of the memory preserving behavior also persists in more irregular networks. This finding opens the possibility to generalize this effect in a wider range of network topology.  

Together with the memory-preserving effect we have also analyzed the spreading of a single excitations in different networks. We have shown that due to cooperative shielding, counter-intuitively, the more connected is the network or the more long ranged are the couplings, the less the initial excitation spreads through the network. 
Clearly, a lower excitation spreading can be obtained also in very low connected network, see Fig.~(\ref{sw_ipr_merit}). 
These findings points out to a possible optimal connectivity for the excitation mobility, or even the possible existence of a sort of percolation transition driven by the connectivity. This topic will be explored in a future work. 

Finally, new evidence of the cooperative shielding effects have been shown in power-law connected networks, where due to cooperative shielding a strong dependence on the initial conditions and on the range of the interaction have been shown, see Fig~(\ref{fig:ipr-powerlaw}).  
  
The results presented in this manuscript open different new paths in the study of highly connected networks.

\begin{acknowledgments}
We would like to thank Concetta Morrone and Marco Cicchini for decisive discussions on the neuroscience application of the analytical findings of this work, specifically concerning the connection between the number of frequencies and the number of excitations, and of the energy bandwith scaling with the Weber's law. S.A. gratefully thanks École polytechnique and Karyn Le Hur, together with the Quantum Sensing Group, Samuel Lellouch and Michael Holynski, for allowing the fulfillment of the present research. J.Y.M. and M.L.C. were supported by the European Social Fund REACT EU through the Italian national program PON 2014-2020, DM MUR 1062/2021. M.L.C. acknowledges support from the National Centre on HPC, Big Data and Quantum Computing—SPOKE 10 (Quantum Computing) and received funding from the European Union Next-GenerationEU—National Recovery and Resilience Plan (NRRP)—MISSION 4 COMPONENT 2, INVESTMENT N. 1.4—CUP N. I53C22000690001. This research has received funding from the European Union’s Digital Europe Programme DIGIQ under Grant Agreement No. 101084035. M.L.C. also acknowledges support from the project PRA2022202398 “IMAGINATION.”
M.L.C, F.B. and G.L.C. are grateful to the KITP for the valuable hospitality during the conception of part of this work. This research was supported in part by grant NSF PHY-2309135 to the Kavli Institute for Theoretical Physics (KITP).
\end{acknowledgments}

\appendix

\section{Degeneracy formula for the XX, all-to-all connected energy levels}
\label{appdx:degeneracy_formula}
In Section~\ref{sec:symm_and_dyn} we introduced the recursive formula $f(l,L)$ for computing the degeneracy of a certain value of the $\op{M}^2$ quantum number $l$ for a given system size $L$; here we prove the result.
As the Hamiltonian in Eq.\eqref{hamiltonian30} is written in terms of global spin operators, the Hilbert space of possible spin configurations (Pauli strings) can be decomposed into irreducible representations of spin $l$, each one with dimension $2l+1$: this means that the total Hilbert space is a direct sum of such representations.

\begin{figure}[!tb]
    \centering    \includegraphics[width=0.65\linewidth]{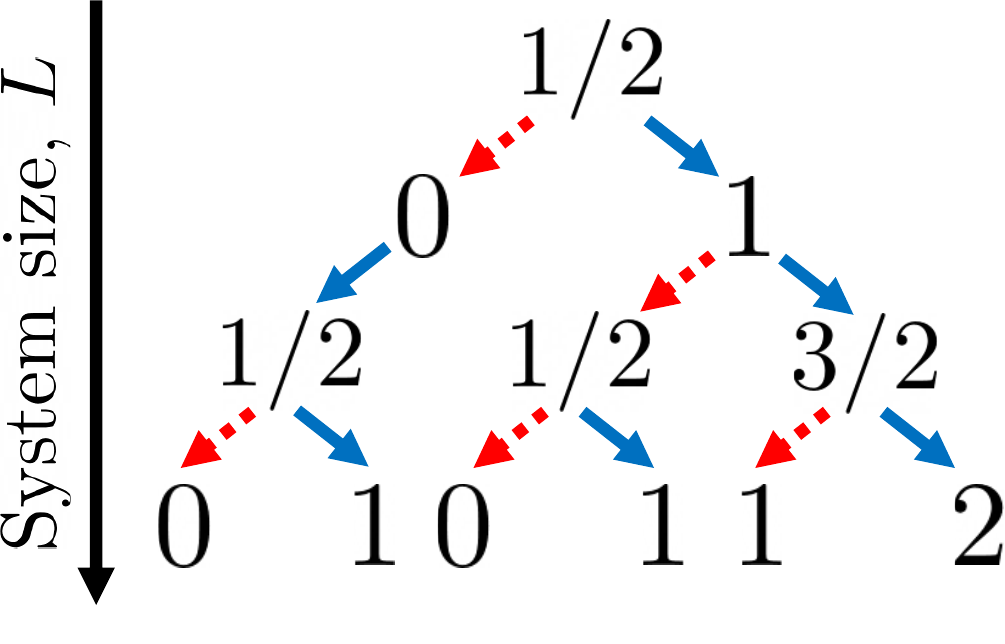}    \caption{Consecutive composition of spin-1/2 angular momenta. The rows of the triangle represent increasing system sizes from 1 to 4. The nodes are the possible values of $l$, obtained either by adding (blue, continuous arrows) or subtracting (red, dashed arrows) 1/2 from the element above.}
    \label{fig:deg_construction}
\end{figure}

Therefore, the question is to calculate the multiplicity of the spin-$l$ representation, given by $f(l,L)$. We build our understanding by induction starting from a single spin-1/2 (the base case): there is only one possible value for $l$, namely 1/2, thus $f(1/2,1)=1$ and any other value of $l$ gives 0. Suppose now that we have system size $L-1$ and we add a new spin-1/2: the theory of composition of angular momenta tells us that each original spin-$l'$ from the $L-1$ system will give result to two new subspaces, one with spin-$l=l'+1/2$ and the other with spin-$l=l'-1/2$ (the latter only if $l'\neq 0$). In other words, $f(l,L)$ is the sum of multiplicities of spin-$l-1/2$ and spin-$l+1/2$ representations at the step $L-1$, as expressed by the recursive formula $f(l,L)=f(l-1/2,L-1)+f(l+1/2,L-1)$. A visual construction is given in Fig.~\ref{fig:deg_construction} for system sizes from 1 to 4: at each row we count how many times the allowed value for $l$ appears, which accounts for the spin-$l$ multiplicity. For example, in the case $L=4$, the allowed value for $l$ are 0, 1, 2, so that the respective multiplicities $f(l,L)$ are 2, 3, 1. Once the spin-$l$ multiplicity known, the actual level degeneracy, dependent on $l$ and $|m|$, can be computed as done in the text.

\section{Numerical methods}
Throughout this work we simulated the systems here reported using Exact Diagonalization (ED) techniques: while the errors contained in these methods are limited to approximations made in the algorithms employed (namely Krylov subspace methods to implement the unitary time-evolution), their main disadvantage relies on the exponential scaling of computational resources with the chain size. There are numerical variational strategies based on Matrix Product States (MPS) and Density Matrix Renormalization Group (DMRG) \cite{schollwoeck2011mps} that have met great success in simulating strongly-correlated quantum systems. However, we employed ED for simulating the quantum dynamics of both all-to-all and regular networks because of the exponential complexity in the number of time steps associated with MPS/DMRG methods.

We are able to circumvent the problem of system size exponential complexity of the ED method thanks to the Hilbert space restriction to symmetry sectors. Indeed, through the computational basis introduced in Section~\ref{sec:symm_and_dyn}, the Hamiltonian~\eqref{hamiltonian1} can be block-diagonalized, each block being defined by the set of eigenstates with a given total $z$-magnetization of the state. This symmetry implies that if we initially consider a state with a well-defined magnetization, our computation of the dynamics simplifies enormously, as it reduces the size of the operators' matrices that we need to compute: for example, by introducing initial states possessing an initial magnetization of $m=(-L/2)+1$ (meaning all spins down except for one) the subspace of relevance will be of size $L$ (binomial coefficient “$L$ choose 1”) instead of $2^L$.

\section{Inverse Participation Ratio}
\label{appdx:ipr_def}
We provide here a compact description of the Inverse Participation Ratio (IPR), used as a quantifier of the (de)localization of an initial excitation. Given a density matrix $\op{\varrho}(t)$ and the subspace of the $L$-dimensional computational basis $\mathcal{P}_1$ describing single-excitation localized states $\{|s\rangle\}$, the IPR is then the sum of $\op{\varrho}$ trace elements squared:
\begin{equation*}
    \text{IPR}(t)=\sum_s \langle s|\op{\varrho}(t)|s\rangle^2\,.
\end{equation*}
In the case where $\op{\varrho}$ is a pure state, the IPR assumes the form:
\begin{gather*}
    |\psi(t)\rangle=\sum_s a_s(t)|s\rangle\quad\Longrightarrow\quad
    \text{IPR}(t)=\sum_s |a_s(t)|^4\,.
\end{gather*}

The IPR is always comprised between $1/L$ and 1, corresponding to the limits of complete delocalization, i.e. infinite temperature case $\op{\varrho}=\op{\mathbb{I}}/L$, and perfect localization of the excitation in a state of our eigenbasis $|s_*\rangle$, respectively. Put it another way, the Participation Ratio (PR), which is the IPR's inverse, indicates how many elements of the chosen basis “significantly” participate to a pure state's temporal evolution, the lower meaning higher localization.

\section{Characterizing complex graphs: average distance and clustering}
\label{appdx:avg_dist_clust}
We provide here the main definitions and relations for complex-networks properties quantifiers.

We first introduce the topological distance dist$_{ij}$ between vertices $i$ and $j$ as the number of edges contained in their shortest path. The average distance $\mathcal{D}$ is its average over all possible couples:
\begin{equation*}
    \mathcal{D}=\frac{1}{\frac{L(L-1)}{2}}\sum_{i<j}\text{dist}_{ij}\,.
\end{equation*}

The clustering coefficient is instead a measure of the nodes' tendency to congregate. While there are many possible definitions, we refer to the one given by Watts and Strogatz in \cite{watts1998smallworld} since it is defined for the class of graphs we consider in this work, i.e. small-world networks. Specifically, it refers to the global clustering coefficient computed as average of the local ones. Given a node $q$ with a number of neighbors given by the degree deg$_q$, we define the local clustering coefficient $C_q$ as the proportion $t_q$ of the number of triangles, i.e. the closed loops consisting of 3 nodes, containing $q$, counted with respect to the maximum number of allowed connections between node $q$ and its neighbors, $\text{deg}_q(\text{deg}_q-1)/2$ (if deg$_q$ is greater than 1): the latter is also referred in the literature as the overall amount of triplets, or subgraphs composed of three nodes connected with at least two edges. Using the properties of the graph's adjacency matrix $A$ ($A_{ij}=1$ if vertices $i$ and $j$ are connected, 0 otherwise), specifically the one that says that its powers indicate the number of walks between two sites, one can compute $t_q$ as the matrix element $(A^3)_{qq}$ divided by 2 (graphs are undirected, so this allows to count the unique triangles). In summary, the clustering coefficient $\mathcal{C}$ comes as follows:
\begin{eqnarray*}
    t_q && =\frac{(A^3)_{qq}}{2} \\
    C_q && =\begin{cases}
        0 \quad \text{if deg$_q$=0, 1} \\
        t_q/[\text{deg}_q(\text{deg}_q-1)/2] \quad \text{if deg$_q$}$>1$
    \end{cases} \\
    \mathcal{C} && =\frac{1}{L}\sum_q C_q\,.
\end{eqnarray*}
An intuition on this peculiar form of the clustering coefficient comes from social networks. Imagine a friendship group: your friends are generally friends between them. Besides, one of them can have a “long-distance connection” with another group. Such a link has the effect to reduce the level of assemblage of that association. \\
Let us make two limiting examples for a nearest-neighbor and completely-connected networks of size $L$. In the former case, $\mathcal{C}$ is 0 since the degree of each node is 2 hence no triangle can be formed, and $\mathcal{D}$ can be shown to scale linearly with $L$ because the sum $\sum_{i<j}\text{dist}_{ij}$ can be written ($L$ even, ring graph) as:
\begin{align*}
    \sum_{i<j}\text{dist}_{ij} & =\underbrace{\Bigl[1+\dots+\frac{L}{2}+\Bigl(\frac{L}{2}-1\Bigr)+\dots+1\Bigr]}_{\text{site 1}} \\
    & +\underbrace{\Bigl[1+\dots+\frac{L}{2}+\Bigl(\frac{L}{2}-1\Bigr)+\dots+2\Bigr]}_{\text{site 2}} \\
    & +\dots \\
    & +\underbrace{[1]}_{\text{site }L}\,, \\
\end{align*}
which grows as $L^3$, while the total number of pairs as $L^2$. In the latter case, both $\mathcal{D}$ and $\mathcal{C}$ are 1, since every vertex is connected to the other by a link and all triplets turn out to be closed, thus becoming triangles.


\bibliography{bibliography}

\end{document}